\documentclass{jfm}
\usepackage{graphicx}
\usepackage{epstopdf,epsfig}
\usepackage{newtxtext}
\usepackage{newtxmath}
\usepackage{natbib}
\usepackage{hyperref}
\hypersetup{
    colorlinks = true,
    urlcolor   = blue,
    citecolor  = black,
}

\newcommand{\RomanNumeralCaps}[1]
\nolinenumbers
\usepackage{amsfonts,amssymb}
\usepackage{longtable} 
\usepackage{multirow} 
\usepackage{amsmath} 
\usepackage[T1]{fontenc} 
\usepackage{tabularx} 
\usepackage{changepage} 
\usepackage{natbib} 
\usepackage{float} 
\usepackage{xcolor} 
\usepackage{relsize} 
\usepackage{url} 
\usepackage{appendix} 
\usepackage{amssymb} 
\newcommand{\sign}{\text{sgn}} 

\title{A scale-wise analysis of intermittent momentum transport in dense canopy flows}

\author{Subharthi Chowdhuri\aff{1}
  \corresp{\email{subharthi.cat@tropmet.res.in}},
  Khaled Ghannam \aff{2}
 \and Tirtha Banerjee\aff{3}}

\affiliation{\aff{1}Indian Institute of Tropical Meteorology, Dr. Homi Bhaba Road, Ministry of Earth Sciences, Pune, 411008, India
\aff{2} Atmospheric and Oceanic Sciences, Cooperative Institute for Modeling the Earth System, Princeton University, Princeton, NJ 08544, USA
\aff{3}Department of Civil and Environmental Engineering, University of California, Irvine, CA, 92697, USA}

\begin{document}
\maketitle

\begin{abstract}
We investigate the intermittent dynamics of momentum transport and its underlying time scales in the near-wall region of the neutrally stratified atmospheric boundary layer in the presence of a vegetation canopy. This is achieved through an empirical analysis of the persistence time scales (periods between successive zero-crossings) of momentum flux events, and their connection to the ejection-sweep cycle. Using high-frequency measurements from the GoAmazon campaign, spanning multiple heights within and above a dense canopy, the analysis suggests that when the persistence time scales ($t_p$) of momentum flux events from four different quadrants are separately normalized by $\Gamma_{w}$ (integral time scale of the vertical velocity), their distributions ($P(t_p/\Gamma_{w})$) remain height-invariant. This result points to a persistent memory imposed by canopy-induced coherent structures, and to their role as an efficient momentum transport mechanism between the canopy airspace and the region immediately above. Moreover, $P(t_p/\Gamma_{w})$ exhibits a power-law scaling at times $t_{p}<\Gamma_{w}$ with an exponential tail appearing for $t_{p} \geq \Gamma_{w}$. By separating the flux events based on $t_p$, we discover that around 80\% of the momentum is transported through the long-lived events ($t_{p} \geq \Gamma_{w}$) at heights immediately above the canopy while the short-lived ones ($t_{p} < \Gamma_{w}$) only contribute marginally ($\approx$ 20\%). To explain the role of instantaneous flux amplitudes towards momentum transport, we compare the measurements with a newly-developed surrogate data and establish that the range of time scales involved with amplitude variations in the fluxes tend to increase as one transitions from within to above the canopy.
\end{abstract}

\section{Introduction}
\label{sec:intro}
Primarily motivated by the requirement of improved parameterization of land-atmosphere exchange in weather and climate models, the characteristics of turbulent transport in vegetation canopies continues to be a subject of inquiry \citep{harman2007simple,bonan2018modeling}. Challenges in this context arise because compared to canonical rough-wall boundary layers, where momentum absorption occurs entirely as friction at the wall, the innate porosity of a vegetation canopy and vertically distributed drag exerted by its foliage on the flow introduce additional characteristic length scales that modulate the turbulence structure \citep{raupach1981turbulence,finnigan2000turbulence}. Within the canopy, fine-scale turbulence is `energized' by wake production in the lee of individual canopy elements, a process that has been modeled as a von K\`{a}rm\`{a}n vortex shedding mechanism \citep{poggi2009flume, ghannam2015spatio, ghannam2020inverse}. The drag discontinuity at the canopy top results in an inflection point in the mean velocity profile, potentially inducing Kelvin-Helmholtz instabilities that penetrate the canopy volume, as in a mixing layer \citep{poggi2004effect, finnigan2000turbulence}. This superposition of a wide range of eddy scales manifests itself in a highly intermittent velocity field \citep{cava2009effects,keylock2020joint}, encoding the effects of both energy-containing coherent structures and smaller scale inertial motion. The scale-wise dynamics of the intermittent statistics of this velocity field within and above the canopy, and its role in momentum transport are a main focus of the work here.  

Time series analysis of turbulent fluctuations in laboratory and numerical experiments has generally focused on the scaling laws of velocity differences (e.g. structure functions), dissipation rates, and their anomalous behavior at small scales (pertaining to inertial subrange and dissipation scales) in homogeneous and isotropic turbulence \citep{sreenivasan1985fine,kraichnan1990models, frisch1995turbulence,sreenivasan1997phenomenology}. In the context of inhomogeneous and non-Gaussian canopy turbulence, two main aspects of the flow have become especially relevant: i) large amplitude variability in turbulent fluctuations and the underlying ejection-sweep cycle \citep{watanabe2004large,wallace2016quadrant}; and ii) local frequency oscillations (zero-crossing properties) to identify the tendency of turbulence events to cluster together \citep{sreenivasan2006clustering,cava2012role}. The former relies on quadrant analysis and conditional sampling of turbulent velocity fluctuations to disentangle the role of canopy-induced organized structures on the ejection-sweep cycle that modulates the momentum flux, $\overline{u^{\prime} w^{\prime}}$ \citep[e.g.][]{raupach1981turbulence,gao1992conditional,thomas2007flux,cava2009effects,chamecki2013persistence}. Here, $u^\prime$ and $w^\prime$ are the longitudinal (streamwise) and vertical (wall-normal) velocity fluctuations around their local time average (denoted by an overbar). Time series analysis of measurements from a large corpus of field campaigns and laboratory experiments showed that sweeping motion, \textit{viz.} downbursts associated with high horizontal momentum ($u^\prime > 0$) swept into the canopy ($w^\prime < 0$), dominate momentum transport. However, these sweeping events are short-lived and therefore lead to intense amplitude fluctuations \citep{thomas2007flux,chamecki2013persistence}, compared to quiescent and low-amplitude ejections ($u^\prime < 0$ and $w^\prime > 0$). This asymmetry between sweeps and ejections is tied to canopy-induced coherent structures in the flow \citep[e.g.][]{finnigan2000turbulence,patton2016atmospheric,cava2012role}, and often is used to explain the nonlocal (or counter-gradient) nature of turbulent fluxes in canopies \citep[e.g.][]{poggi2004momentum,chamecki2020effects}.

A complementary aspect of time series analysis aims at isolating the amplitude fluctuations in turbulent signals using zero-crossing properties, called persistence analysis \citep{bray2013persistence,ghannam2016persistence,chowdhuri2020persistence_a}. Borrowed from the study of spatially-extended nonequilibrium systems \citep{majumdar1999persistence}, persistence represents the probability that a stochastic process (e.g. here $u^{\prime}$, $w^{\prime}$, or even $u^{\prime}w^{\prime}$) remains in a certain state (below/above a prescribed threshold) up to some time $t_{p}$. As turbulent velocity fluctuations are driven by a spectrum of scales, spanning a wide range of coherent structures, they tend to exit and re-enter such states as time evolves. The time scale $t_{p}$ therefore becomes a random variable with a probability density function $P(t_p)$, the scaling of which is indicative of the nature of the underlying dynamics \citep{sreenivasan2006clustering}. 

Earlier theoretical and experimental studies for many dynamical systems showed that the persistence probability decays as a power law at late times, although this scaling is usually nontrivial \citep{majumdar1999persistence}. For instance, the study by \cite{kailasnath1993zero} suggested a double exponential scaling for $P(t_p)$ in turbulent boundary layers with two distinct time scales, where at long times the time scale becomes independent of the Reynolds number ($Re$). In contrast, a variety of field experiments in high $Re$ flows in the atmospheric surface layer (ASL) and roughness sublayer (RSL) of plant canopies suggest a power law scaling for $P(t_p)$ at short times, and an exponential tail with a cutoff time scale comparable to the turbulence integral time scale \citep{cava2009effects, cava2012role,chamecki2013persistence,chowdhuri2020persistence_a}. Notwithstanding, the connection between these dominant persistence time scales and the flow structure, often aided by the use of Taylor's frozen turbulence hypothesis \citep{taylor1938spectrum}, remains an outstanding topic. This is particularly the case in the context of RSL turbulence, where the role of canopy-induced coherent structures in the flux generation mechanism is of paramount significance.

In this study, we revisit the connections between the turbulence structure in the vicinity of vegetation canopies and the persistence in measured velocity fluctuations that underlie the mechanisms of momentum flux transport. The central hypothesis rests on the conditional sampling of the velocity field to disentangle the role of low-amplitude but persistent ejections, and intense but short-lived sweeps, in shaping the dominant time scales of the flow relative to the integral scales. More specifically, we aim to answer the following questions: (1) Can the scaling properties of persistence behavior be related to the presence of coherent turbulence structures in a canopy flow? (2) How important is the role of instantaneous flux amplitudes associated with the $u^{\prime}w^{\prime}$ events of different persistence time scales while explaining the momentum transport in a canopy flow? (3) Does there exist any range of event time scales for which the amplitude variability in $u^{\prime}w^{\prime}$ contribute most to the momentum transport within and above the canopy heights?

To carry out our research objectives, sonic anemometer measurements from a dense vegetation canopy in the Amazon forest are used, and these span multiple heights within and above canopy to investigate the role of canopy drag on the distribution of persistence times. The details of this dataset and the processing steps are outlined in \S\,\ref{data_methods}. In order to delineate the mechanisms of intermittent momentum transport, persistence analysis is employed on the dataset and the obtained results are presented in \S\,\ref{results}. Finally, in \S\,\ref{conclusion} conclusions are drawn and future research directions are provided.

\section{Dataset and methodology}
\label{data_methods}
\subsection{Field data}
\label{data}
We use the observational dataset from the GoAmazon experiment \citep{fuentes2016linking,freire2017turbulent,gerken2018investigating, ghannam2018scaling}. The data were collected during a field campaign at the Cuieiras Biological Reserve, located $\approx$ 60 km North-Northwest of the city of Manaus, Brazil, between March 2014 and January 2015 at a 50-m-tall tower surrounded by a dense primary forest. The average canopy height at the measurement site is $h \approx$ 35 m. The leaf area index (LAI), which is defined as the total one-sided leaf area (half the total foliage area) per unit ground surface area, was estimated to be between 6.1 and 7.3 m$^{2}$ m$^{-2}$ \citep{fuentes2016linking}. The atmospheric boundary layer depth $\delta$, especially over a tropical area like the Amazon, can be on the order of 1000 to 2000 m, such that the friction Reynolds number $Re=(u_{*}\delta)/\nu \approx 10^{7}$, where $u_{*}$ is the friction velocity at the canopy top and $\nu$ is the kinematic viscosity of air. A more readily available quantity is the local Reynolds number based on the canopy height, viz. $Re_{h}=(u_{*}h)/\nu \approx 10^{5}$. In either definition, the Reynolds number of the flow is extremely large compared to those achieved in laboratory experiments or numerical simulations. 

Regarding the instrumentation part, high frequency time series of the three wind velocity components and sonic temperature within and immediately above the canopy were measured by nine time-synchronized triaxial sonic anemometers (CSAT3, Campbell Scientific, Logan, Utah). The measurement frequency was 20 Hz, and measurement heights were $z/h =\{$0.2, 0.39, 0.52, 0.63, 0.7, 0.9, 1, 1.15, 1.38$\}$, where $z$ is the height referenced to the ground surface. For all the nine sonic anemometers, the data were divided into 30-min blocks and a double-coordinate rotation was applied to align the $x$-axis with the direction of the mean velocity \citep{kaimal1994atmospheric}. Turbulent fluctuations in the wind components ($u^{\prime}$, $v^{\prime}$, and $w^{\prime}$ in the streamwise, cross-stream, and vertical directions respectively) and in the sonic temperature ($T^{\prime}$) were computed by subtracting the mean.

Four additional criteria were also imposed to select data blocks for subsequent analysis: i) mean wind direction at the highest anemometer is within $\pm 90^{\circ}$ from the axis of the anemometer, such that the wind faces the anemometer directly; ii) there exists an inertial subrange following a 2/3 power-law in the second-order structure functions \citep{kolmogorov1941local}; iii) stationarity of the horizontal wind as proposed by \citep{vickers1997quality}; and iv) near-neutral stratification, specified by the condition $|(z-d)/L| \leq 0.5$ at and above the canopy top ($z/h=$ 1, 1.15, and 1.38), where $d$ is the displacement height assumed constant ($d=2h/3$), and $L$ is the Obukhov length. Note that \cite{ghannam2018scaling} estimated the displacement height for this canopy as the centroid of the drag force and found a consistent value of $d \approx 0.67h$. The criterion $|(z-d)/L| \leq 0.5$ for near-neutral stability is similar to that used in \citet{cava2009effects} and \citet{cava2012role}. 

In the present study, a total of 93 data blocks (30-min runs) that meet the above criteria are analyzed, where each block consists of time-synchronized measurements from nine observation heights. For these data blocks, the momentum ($\overline{u^\prime w^\prime}$) and heat ($\overline{w^\prime T^\prime}$) fluxes were fairly constant above the canopy, i.e., corresponding to the heights $z/h \geq 1$. Additionally, the friction velocity at the canopy top ($u_{*}$) was estimated to be equal to 0.46 m s$^{-1}$ after taking an ensemble average over the 93 near-neutral runs, with a run-to-run variation between 0.4 to 0.6 m s$^{-1}$. This ensemble mean of $u_{*}$ is typical of canopy turbulence under near-neutral conditions \citep[e.g.,][]{chamecki2013persistence}.  

\subsection{Persistence analysis}
\label{methods}

\begin{figure}
\centering
\hspace*{-0.8in}
\includegraphics[width=1.25\textwidth]{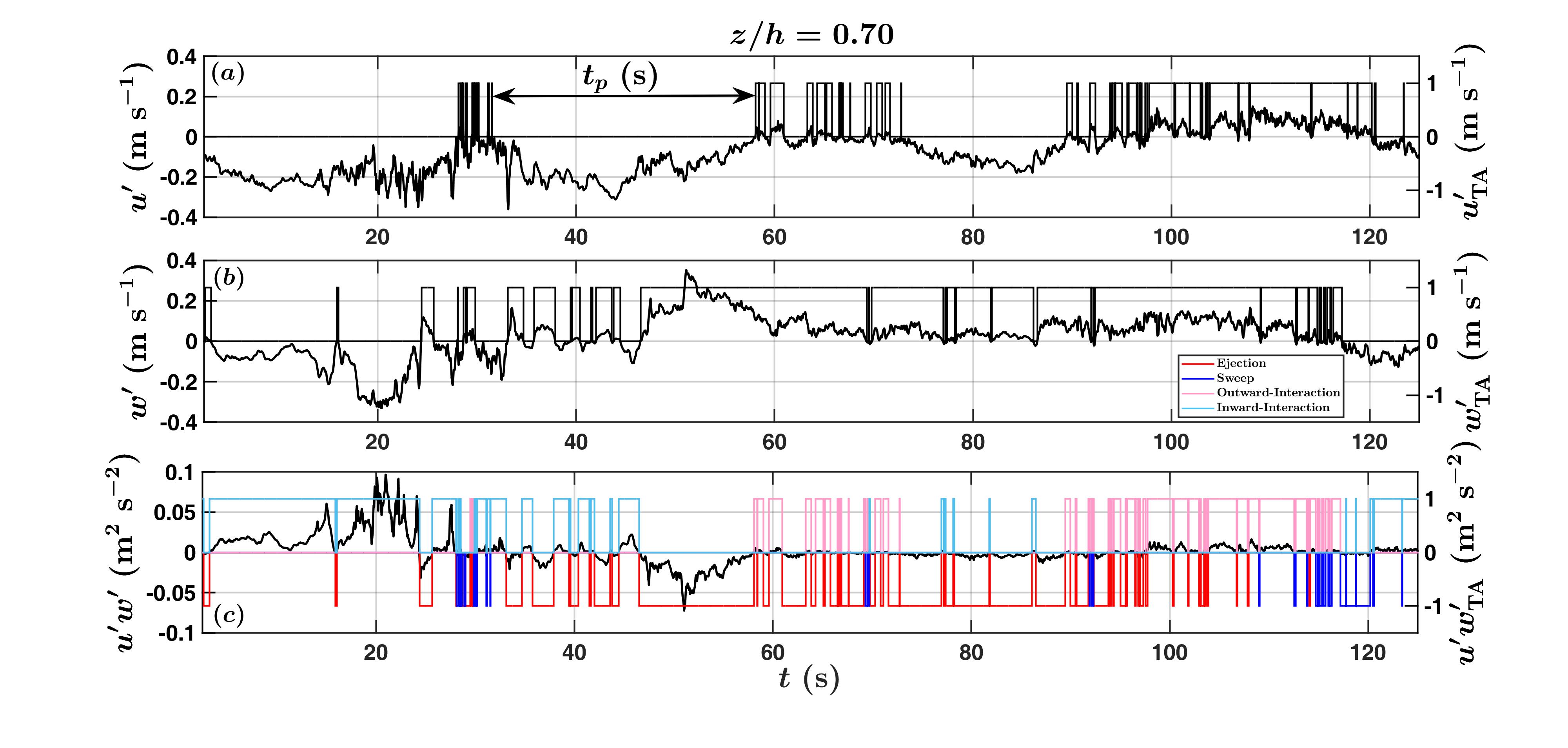}
 \caption{The 20-Hz time-series of (a) $u^{\prime}$, (b) $w^{\prime}$, and (c) $u^{\prime}w^{\prime}$ are shown at height $z/h=0.7$ between the duration 2.5--125 s, to explain the concept of persistence time scale ($t_{p}$). In (a)--(c), the right hand side of the $y$ axis represents the telegraphic approximations (TA) of the original time series. For the $u^{\prime}w^{\prime}$ time-series, four distinct states are possible where the flux values reside, based on the ejection, sweep, outward-interaction, and inward-interaction quadrants.}
\label{fig:1}
\end{figure}

For these selected near-neutral runs, we apply persistence analysis on the streamwise and vertical velocity fluctuations ($u^{\prime}$ and $w^{\prime}$) and on the instantaneous momentum flux signals ($u^{\prime}w^{\prime}$). In a time-series analysis, persistence is defined as the probability that the local value of a fluctuating field does not change sign (or equivalently its state) up to a certain time $t$ \citep[e.g.,][]{bray2013persistence,chowdhuri2020persistence_a,chowdhuri2019revisiting}. Therefore, the foundation of persistence analysis is completely rooted in the temporal domain unlike the Fourier or wavelet methods, allowing one to directly define the time scales from the switching properties of the signal rather than employing any mathematical functions. Moreover, as discussed in \citet{platt1993off}, the abrupt or aperiodic switching between qualitatively different kind of states is related to an aspect called on-off intermittency in the parlance of fluid turbulence. 

By combining such concept with the definition of persistence, one can infer that persistence analysis offers the unique flexibility to study the time scales associated with on-off intermittency. Eventually, this inference boils down to studying separately the time scales of positive and negative fluctuations in any turbulent signals or the ones linked to different transport modes of turbulent fluxes (occurring aperiodically). Depending on the context, persistence is also referred to as distributions of the first-passage time, or survival probability distributions, or return-time distributions, or the distributions of the inter-arrival times between the successive zero-crossings \citep{laurson2009effect,castellanos2013intermittency,grebenkov2020preface,kumar2020interevent}. 

A graphical demonstration of persistence is provided in figure \ref{fig:1}a-b, where nearly 120-s long sections of $u^{\prime}$ and $w^{\prime}$ signals are shown for a particular 30-min time series from near-neutral conditions, at height $z/h=0.70$. From figure \ref{fig:1}a, one can see that the $u^{\prime}$ signal displays persistent positive or negative values (with respect to the mean) for a particular amount of time, denoted as $t_{p}$. Since $t_{p}$ can also be interpreted as the inter-arrival time between the subsequent zero-crossings, those are identified by using the telegraphic approximation (TA) as,
\begin{equation}
(x^{\prime})_{\rm TA}=\frac{1}{2}(\frac{x^{\prime}(t)}{|x^{\prime}(t)|}+1), \ x=u,w
\label{TA}
\end{equation}
and locating the points where the TA series changes its value from 0 to 1 or vice-versa (see the right-hand-side axes of figure \ref{fig:1}a-b). More importantly, the interaction between the persistence patterns in $u^{\prime}$ and $w^{\prime}$ signals give rise to the time scales of the $u^{\prime}w^{\prime}$ quadrant cycles, as shown in figure \ref{fig:1}c. Depending on when the $u^{\prime}$ and $w^{\prime}$ signals switch from 0 to 1 or vice-versa, the residence times in four quadrants are decided. The four quadrants can be identified in figure \ref{fig:1}c through the bar-plots of four different colours (see the legend in figure \ref{fig:1}b), where $+$1 is used for counter-gradient quadrants and $-$1 for co-gradient quadrants (see the right-hand-side axis of figure \ref{fig:1}c). In table \ref{tab:1} of appendix \ref{app_B} one can find the descriptions associated with four different quadrants of $u^{\prime}w^{\prime}$ signal. 

\subsubsection{Probability density functions}
\label{persistence_PDFs}
To characterize the statistical properties of the time scales, we compute the probability density functions (PDFs) of $t_{p}$ ($P(t_{p})$) or the persistence PDFs, corresponding to the $u^{\prime}$, $w^{\prime}$, and $u^{\prime}w^{\prime}$ signals. For computing these PDFs, standard statistical procedures are adopted, which involve logarithmic binning (see \citet{chowdhuri2020persistence_a} for a detailed explanation). By combining all the 93 near-neutral runs, we encounter more than 10$^{5}$ number of zero-crossing events for each signal. The persistence PDFs are constructed over these large number of event ensembles, and therefore, remain statistically quite robust. Moreover, in our computations of $P(t_{p})$ we use 60 logarithmic bins of $t_{p}$ by following \citet{chowdhuri2020persistence_a}. To get an idea about how many events are associated with each bin, we evaluated the histograms of $t_{p}$ for $u^{\prime}$, $w^{\prime}$, and $u^{\prime}w^{\prime}$ signals (not shown). From such histograms we could notice a preponderance in the number of events for the bins that represented short time scales, and a number significantly larger than 100 for events corresponding to the larger time scales. After obtaining the persistence PDFs, we next investigated how much each persistence patterns from one of the four quadrants of $u^{\prime}$-$w^{\prime}$ contributed to the momentum flux.

\subsubsection{Flux contributions}
\label{flux_contributions}
For estimating the quadrant contributions to the momentum flux against the persistence time scales, one first identifies the persistence patterns from any of the four quadrants (see figure \ref{fig:1}c), estimates their time scales ($t_{p}$), and then computes the amount of instantaneous momentum flux ($u^{\prime}w^{\prime}$) carried within such patterns. Following \citet{chowdhuri2019revisiting}, this operation can be mathematically expressed as,
\begin{align}
\begin{split}
\Big \langle u^{\prime}w^{\prime}\vert\Big[t_{p}\{\rm bin\}<t_{p}< t_{p}\{\rm bin\}+d\log(t_{p})\Big] \Big \rangle &=
\\
\frac{\sum{u^{\prime}w^{\prime}}}{N \times d\log(t_{p})},
\end{split}
\label{F_1}
\end{align}
where $t_{p}\{\rm bin\}$ is the logarithmically binned value of $t_{p}$, $d\log(t_{p})$ is the bin-width, and $N$ is the number of samples in a 30-min run. The division by $N$ and $d\log(t_{p})$ is performed in Eq. (\ref{F_1}) to ensure that when integrated over all the possible $t_{p}$ values, the result would be the momentum flux contribution from a particular $u^{\prime}$-$w^{\prime}$ quadrant. Moreover, since $d\log(t_{p})=dt_{p}/t_{p}$ and is used in the denominator, the flux distribution in Eq. (\ref{F_1}) can be considered as a premultiplied one with $t_{p}$. Henceforth, to be more concise, the quantity at the left hand side of Eq. (\ref{F_1}) is simply denoted as $\langle u^{\prime}w^{\prime}\vert(t_{p})\rangle$. Note that, in figure \ref{fig:5}e--h where we show this quantity, a suitable normalization scheme is employed to non-dimensionalize both the persistence time scales and their associated flux values.   

The plot of the quantity $\langle u^{\prime}w^{\prime}\vert(t_{p})\rangle$ against $t_{p}$, is equivalent to what is generally referred as size-duration scaling relation in the context of avalanche dynamics where one expects the avalanche sizes to scale with the duration as a power-law \citep{laurson2009effect,castellanos2013intermittency,benzi2022self}. The avalanche sizes are basically considered as the integrated quantities of signal values up to a certain duration $t_{p}$, and therefore, share a perfect resemblance with our momentum flux distributions defined in Eq. (\ref{F_1}). Some previous studies have analytically shown that for a Brownian motion, the size-duration scaling relationship is a power-law with an exponent $1.5$ \citep{laurson2009effect}. By exploiting this fact, \citet{castellanos2013intermittency} identified specific regions in a turbulent plasma flow, which were governed by non-diffusive transport mechanisms.

Apart from the connection with Brownian motions, these plots also provide an idea whether accounting for only the zero-crossing or persistence properties is enough when one contemplates flux modelling. For instance, let us assume a flux event from any one of the quadrants that persists for time $t_{p}$ before switching its state. Additionally, consider the point that the amplitude variability is negligible for this particular event, i.e., the instantaneous $u^{\prime}w^{\prime}$ values are nearly constant over that duration. In such case the flux carried within that event should be directly proportional to its time scale $t_{p}$. Further, if this is true for all the events of different persistence time scales, then the flux contributions can be predicted from the persistence PDFs alone. Therefore, under these circumstances, one may expect the flux contributions would scale similarly as the persistence PDFs. On this note, \citet{laurson2009effect} analytically showed that if the persistence PDFs follow a power-law with an exponent $-\alpha$, then the size-duration scaling would exhibit an exponent $+\alpha$ if the amplitude variations are absent. More discussions on these aspects are presented in \S\,\ref{flux_contr}.

\section{Results and discussion}
\label{results}
We begin by discussing the vertical profiles of bulk flow statistics and properties of momentum transport within and above the canopy. To gain insights on the time scales of turbulence structures associated with momentum transport, persistence analysis is employed. Based on the statistical characteristics of such time scales, two distinct event classes (indicative of certain flow structures) are recognized and their contributions to the momentum flux are investigated separately. We conclude by dissecting the role of $u^{\prime}w^{\prime}$ amplitude variability from persistence effects, in order to explain momentum transport associated with the flux events in canopy flows. 

\subsection{Statistical description of the flow}
\label{stats}
\subsubsection{Vertical profiles}
\label{vert_prof}
\begin{figure}
\centering
\hspace*{-1in}
\includegraphics[width=1.4\textwidth]{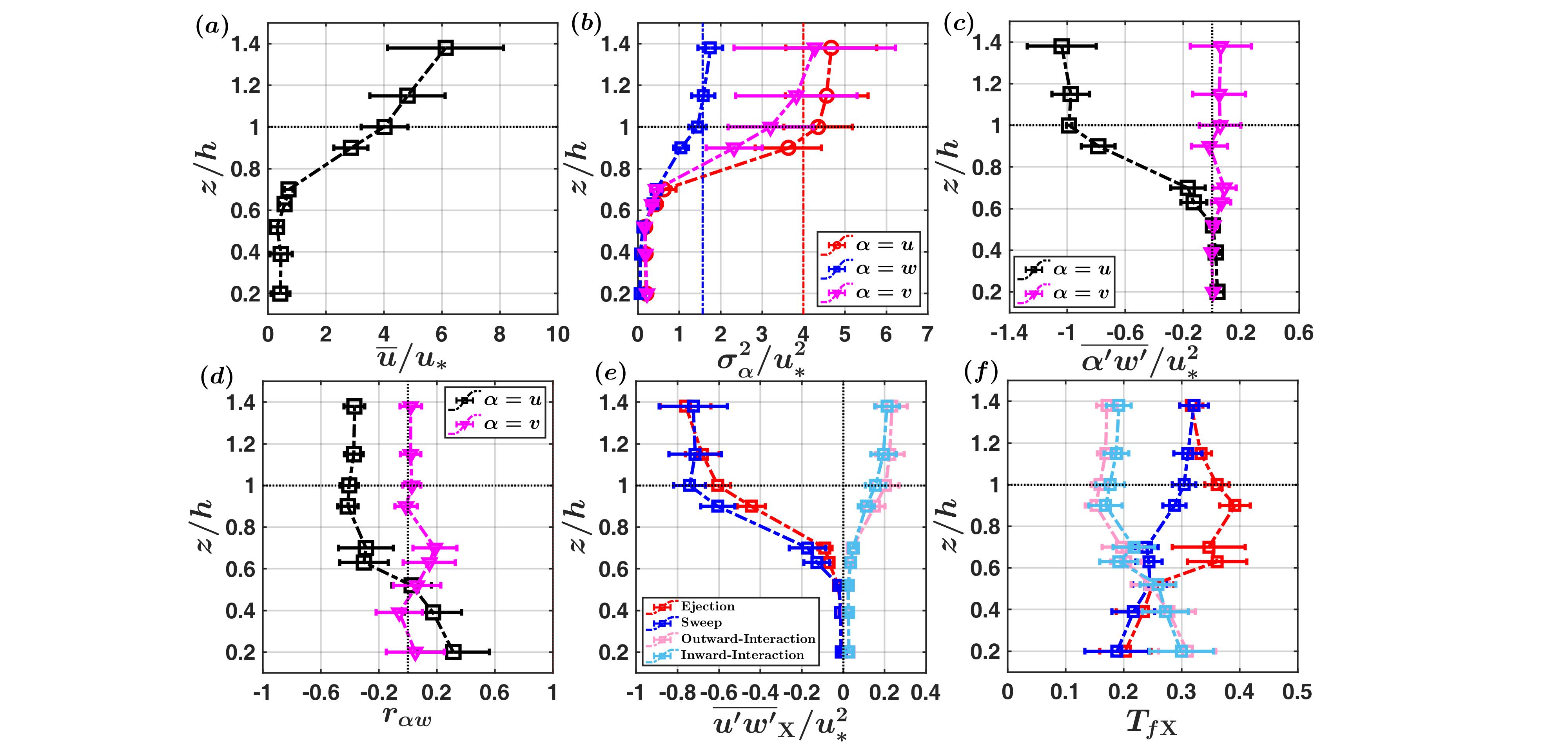}
 \caption{Ensemble averaged profiles of (a) mean velocity ($\overline{u}/u_{*}$), (b) velocity variances ($\sigma^{2}_{\alpha}/u^{2}_{*}$, where $\alpha=$ $u$, $v$, or $w$), (c) momentum fluxes ($\overline{\alpha^{\prime}w^{\prime}}/u^{2}_{*}$, where $\alpha=u$ or $v$), (d) correlation coefficients ($r_{\alpha w}$), (e) flux fractions ($\overline{u^{\prime}w^{\prime}}_{\rm X}/u^{2}_{*}$), and (f) time fractions ($T_{f\rm X}$) corresponding to each of the four $u^{\prime}$-$w^{\prime}$ quadrants ($\rm{X}$ denotes any one of the four quadrants). The ordinate axis is the normalized height $z/h$, and $u_{*}$ is the friction velocity at the canopy top. The error-bars in all the panels denote one standard deviation from the averaged values.}
\label{fig:2}
\end{figure}

Figure \ref{fig:2} shows the vertical profiles of some flow statistics plotted against the normalized height ($z/h$). Unless otherwise specified, the statistical quantities in figure \ref{fig:2} and in the rest of the article are ensemble averages over the selected 93 data blocks from near-neutral conditions, and the velocity scale $u_{*}$ is the friction velocity computed at the canopy top. To denote the spread, error-bars in figure \ref{fig:2} represent one standard deviation from the ensemble mean.

Overall, the statistics depicted in figure \ref{fig:2} are in agreement with field observations and large-eddy simulation experiments reported in previous studies \citep[e.g.][]{finnigan2000turbulence,dias2015large,patton2016atmospheric,banerjee2017connecting,brunet2020turbulent}. The normalized profile of the mean wind speed ($\overline{u}/u_{*}$) displays a sharp gradient in the upper part of this dense canopy, and attains a value of $\overline{u}/u_{*}=$ 4 at $z/h=1$ with increasing up to 6 at $z/h=1.38$ in the roughness sublayer (figure \ref{fig:2}a). The streamwise, cross-stream, and vertical velocity variances ($\sigma_{u,v,w}^2/u^2_{*}$) also remain close to zero at heights deep within the canopy but approach the near-neutral values of $\sigma_{u,v}/u_{*} \approx 2$ and $\sigma_{w}/u_{*} \approx 1.25$ \citep{finnigan2000turbulence} at heights $z/h \geq $ 1 (figure \ref{fig:2}b). Interestingly, the cross-stream velocity variances ($\sigma_{v}^2/u^2_{*}$) agree well with the near-neutral limit while a small deviation is observed for the streamwise ones ($\sigma_{u}^2/u^2_{*}$). 

The strong shear in $\overline{u}$ is reflected in the streamwise momentum flux ($\overline{u^{\prime}w^{\prime}}/u^{2}_{*}$) profile (figure \ref{fig:2}c), where $\approx$ 90\% of momentum is absorbed in the upper 30\% of the canopy, with a constant value atop ($|\overline{u^{\prime}w^{\prime}}| \to u^{2}_{*}$ at $z/h \geq $ 1). On the other hand, the cross-stream momentum fluxes ($\overline{v^{\prime}w^{\prime}}$) remain nearly zero throughout the canopy heights. This suggests, for our near-neutral canopy flow setup, the shear stress vector aligns well with the mean wind direction \citep{pan2017determining}. If one considers the correlation coefficients ($r_{uw}$ and $r_{vw}$), $r_{uw}$ values remain substantially larger than $r_{vw}$ while being slightly positive at the lowest three levels (figure \ref{fig:2}d). The slight positive values of $r_{uw}$ at lowest canopy heights agree well with the observations of \citet{dupont2012influence}. Moreover, at heights $z/h \geq 1$, $r_{uw}$ approximately approaches $-$ 0.4, in agreement with \citet{finnigan2000turbulence} and \citet{dupont2012influence}. Hereafter, while referring to the momentum flux we only imply $u^{\prime}w^{\prime}$.  

Although in subsequent sections more information is provided about $u^{\prime}w^{\prime}$, the simplest way to start is by investigating the quadrant contributions to the time-averaged momentum flux values and how much time does the flow spend in each quadrant (see table \ref{tab:1} explaining the four different quadrants). To compute the amount of flux carried by each quadrant and the fraction of time spent in each, we use,
\begin{align}
\begin{split}
\frac{\overline{u^{\prime}w^{\prime}}_{\rm X}}{u^{2}_{*}}=\frac{\mathlarger{\sum{\Big[(u^{\prime}w^{\prime}})I_{\rm X}\Big]}}{Nu^{2}_{*}},
\\
T_{f\rm X}=\frac{\sum{I_{\rm X}}}{N}, \ (\rm X=E,S,OI,II)
\end{split}
\label{fraction}
\end{align}
where,
\[
I_{\rm X} = \begin{cases}
1 &\text{if $\{u^{\prime},w^{\prime}\} \in \rm{X}$}\\
0 &\text{otherwise}
\end{cases}
\]
and $N$ is the total number of points in a 30-min run (36000 in our case). The results related to these aspects are presented in figure \ref{fig:2}e--f. By evaluating figure \ref{fig:2}e--f, it can be inferred that the differences in characteristics between the ejection and sweep motions remain most prominent for the mid-canopy heights, such as: $z/h=$ 0.63, 0.70, and 0.90. However, as $z/h$ approaches 1 and above, this distinction becomes hardly noticeable. Specifically, at heights $z/h=$ 0.63 to 0.90, flux contributions from the sweep quadrant remain larger than ejection (figure \ref{fig:2}e), despite the fact that the flow spends more time in the ejection quadrant (figure \ref{fig:2}f). Therefore, at those heights, the short-lived sweeps appear to be a little intense than the long-lived ejections. 

\begin{figure}
\centering
\hspace*{-0.95in}
\includegraphics[width=1.3\textwidth]{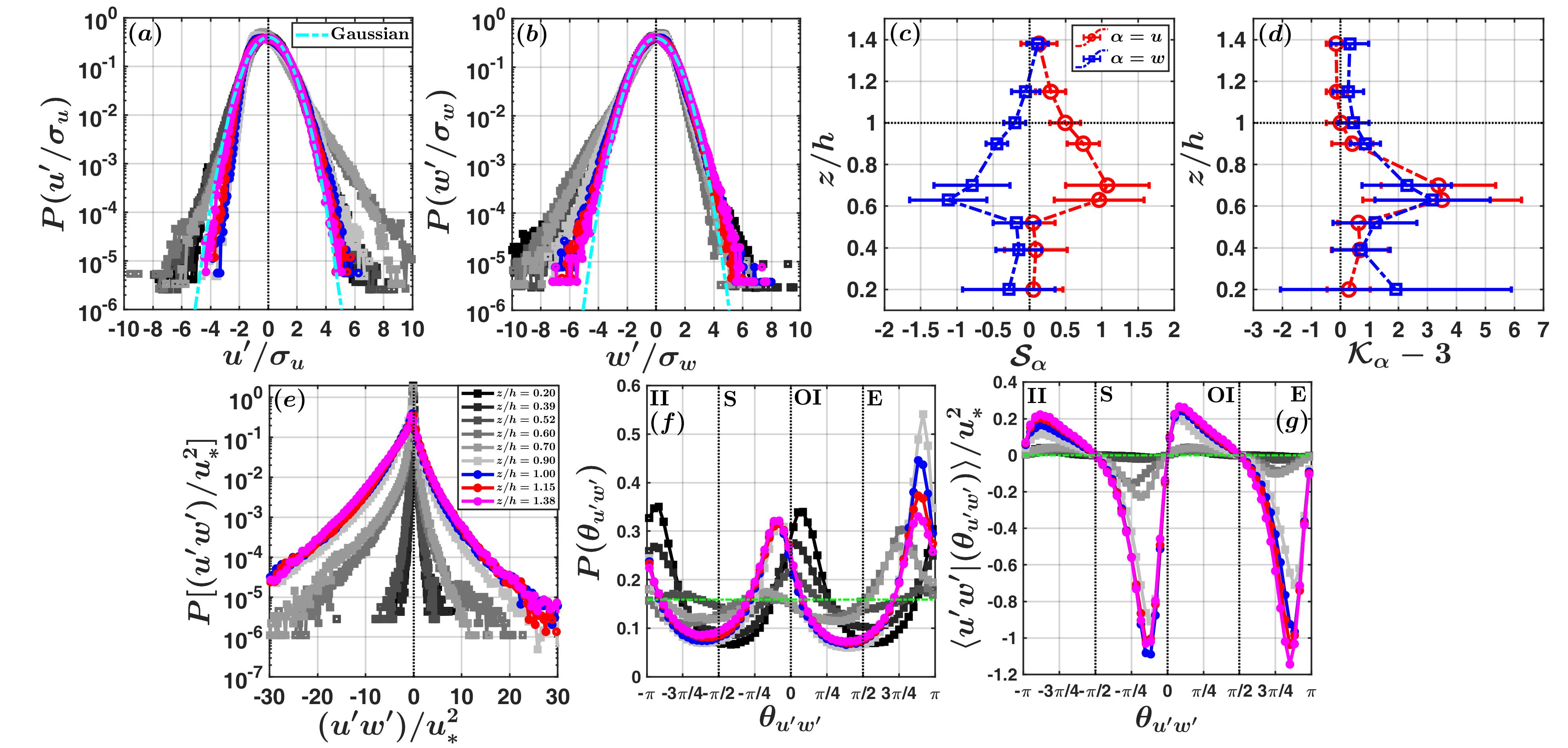}
 \caption{Probability density functions (PDFs) for the normalized (a) streamwise velocity fluctuations ($u^{\prime}/\sigma_{u}$), and (b) vertical velocity fluctuations ($w^{\prime}/\sigma_{u}$) are shown. The standard Gaussian distribution is overlaid on (a) and (b) as cyan dash-dotted lines. In (c) and (d), the vertical profiles of skewness ($\mathcal{S}_{u}$, $\mathcal{S}_{w}$) and excess kurtosis ($\mathcal{K}_{u}-3$, $\mathcal{K}_{w}-3$) of $u^{\prime}$ and $w^{\prime}$ signals are shown. The bottom panels demonstrate (d) PDFs of instantaneous momentum fluxes ($u^{\prime}w^{\prime}/u^{2}_{*}$), (e) PDFs of quadrant angles $\theta_{u^{\prime}w^{\prime}}$, and (f) quadrant distributions of $u^{\prime}w^{\prime}$. The descriptions of different markers for the heights within (gray coloured squares) and above the canopies (blue, red, and pink coloured circles) are depicted in the legend of (d). The dash-dotted green horizontal line in (e) denotes the PDF of a uniform distribution.}
\label{fig:3}
\end{figure}

\subsubsection{PDFs and quadrant analysis}
\label{PDFs}
To further unravel the statistical character of the velocity fluctuations ($u^{\prime}$ and $w^{\prime}$), we present the PDFs of these quantities in figure \ref{fig:3}a--b. It is worth remarking that in these figures and the ones to be followed, gray-coloured squares of various shades denote the heights within the canopy ($z/h<1$), whereas the heights at and above the canopy ($z/h \geq 1$) are represented by blue-, red-, and pink-coloured circles. Before computing these PDFs, we normalized $u^{\prime}$ and $w^{\prime}$ with their respective standard deviations, so that the results can be compared with a standard Gaussian distribution (see the cyan dash-dotted lines in figure \ref{fig:3}a--b). In figure \ref{fig:3}a--b, the PDFs $P(u^{\prime}/\sigma_{u})$ and $P(w^{\prime}/\sigma_{w})$ develop a prominent heavy tail (with respect to a Gaussian) towards the positive and negative side, respectively, at $z/h$ values between 0.63 to 0.90. Other than these heights, the PDFs mostly remain close to Gaussian.

In addition to the PDFs, we present the skewness ($\mathcal{S}_{u,w}$) and kurtosis ($\mathcal{K}_{u,w}$) of $u^{\prime}$ and $w^{\prime}$ signals (figure \ref{fig:3}c--d). Physically, skewness is associated with asymmetry in the PDFs whereas kurtosis is related to intermittency. For a non-intermittent signal, kurtosis is 3, which corresponds to a Gaussian distribution. Therefore, to imply intermittency, instead of showing the kurtosis values directly one can subtract 3 from them (figure \ref{fig:3}d). Regarding skewness, the positive (negative) values denote whether the signal is more dominated by the large positive (negative) fluctuations, i.e., if the PDF has a fat tail towards the positive (negative) side. From both PDFs and skewness plots (figure \ref{fig:3}a--c), it is abundantly clear that at mid canopy heights, the fluctuations in $u^{\prime}$ and $w^{\prime}$ signals are influenced by large positive and negative values, respectively. Because such extreme values are present, $\mathcal{K}_{u,w}$ values at $z/h=$ 0.63, 0.70, and 0.90 remain well above 3, thereby indicating strongly-intermittent signals (figure \ref{fig:3}d). Moreover, for both $u^{\prime}$ and $w^{\prime}$, the maximum $\mathcal{S}_{u,w}$ ($\approx \pm 1$) and $\mathcal{K}_{u,w}$ values ($\approx$ 7) are located at a height $z/h=0.63$. This observation is in agreement with \citet{dupont2012influence}, which they interpreted as an indication of the penetration by the canopy scale coherent structures resulting in strong sweeps and weak ejections at mid canopy heights. Our results from figure \ref{fig:2}e--f support this interpretation.

We next investigate how the height variations in $P(u^{\prime}/\sigma_{u})$ and $P(w^{\prime}/\sigma_{w})$ impact the distributions of the instantaneous momentum flux. Since we are interested in fluxes and not in correlation coefficients, $u^2_{*}$ is used as a scaling factor for the PDFs of $u^{\prime}w^{\prime}$ rather than $\sigma_{u}$ and $\sigma_{w}$ (figure \ref{fig:3}e). While observing figure \ref{fig:3}e, one can notice that $P[(u^{\prime}w^{\prime})/u^2_{*}]$ gradually develop large negative tails with increasing $z$, indicating a more frequent occurrence of the extreme values in $-u^{\prime}w^{\prime}$. This can be further confirmed through the vertical profiles of skewness in $u^{\prime}w^{\prime}$ (not shown). Such plots corroborate the fact that except the lowest three measurement heights, the skewness values are negative, meaning large negative events (ejections and sweeps) dominate the $u^{\prime}w^{\prime}$ signal. Therefore, to better understand the role of ejections and sweeps towards the momentum transport, we introduce polar-quadrant analysis. For a detailed description of this particular methodology, see appendix \ref{app_B} and figure \ref{fig:A2} therein.   

In figure \ref{fig:3}f we show the ensemble-averaged phase angle PDFs $P(\theta_{u^{\prime}w^{\prime}})$, corresponding to all the nine measurement heights. The dotted green line in figure \ref{fig:3}f represents a uniform distribution $P(\theta_{u^{\prime}w^{\prime}})=1/(2\pi)$, whose significance will be discussed in figure \ref{fig:7}a--c. From figure \ref{fig:3}f one can see that for the lowest three heights, the behaviour of $P(\theta_{u^{\prime}w^{\prime}})$ is mainly dominated by the two peaks residing in the outward- and inward-interaction quadrants. But as the heights increase, the peaks in $P(\theta_{u^{\prime}w^{\prime}})$ shift towards the ejection and sweep quadrants. Moreover, at mid canopy heights ($z/h$ between 0.63 to 0.90), the peaks of $P(\theta_{u^{\prime}w^{\prime}})$ remain larger for the ejection quadrant. This is consistent with figure \ref{fig:2}f where one observed at those heights the ejection motions occupied more time with respect to the other three quadrants. Apart from that, at heights $z/h \geq 1$, the sweep and ejection peaks in $P(\theta_{u^{\prime}w^{\prime}})$ are confined to the angles $-\pi/4<\theta_{u^{\prime}w^{\prime}}<0$ and $3\pi/4<\theta_{u^{\prime}w^{\prime}}<\pi$, respectively. This outcome has an important consequence when one considers the flux distributions associated with the phase angles.

The flux distributions (see Eq. (\ref{Flux_theta}) in appendix \ref{app_B}) are shown in figure \ref{fig:3}g where they are normalized by $u^{2}_{*}$ and ensemble averaged over all the near-neutral runs. By attentively observing the flux behaviors at heights $z/h=$ 0.63 to 0.90, one can conclude while the PDF peaks remain larger for the ejections (figure \ref{fig:3}f) the flux contributions are higher for the sweeps (figure \ref{fig:3}g). Note that this apparent distinction between the ejection and sweep quadrants gradually fades as one approaches the heights $z/h \geq 1$ (figure \ref{fig:3}f--g). All such findings are in accordance with figure \ref{fig:2}e--f. Furthermore, from figure \ref{fig:3}f--g, it can be seen that the maximum flux contributions and PDF peaks from the ejection and sweep quadrants are concentrated around those phase angles that lie on the left-hand-side plane of the $-45^{\circ}$ degree slope line (see the dashed black line in figure \ref{fig:A2}). This indicates the events from ejection and sweep quadrants, which contribute most to the momentum flux, are in fact associated with larger magnitudes of $u^{\prime}$ than $w^{\prime}$. Such features appear more prominent at heights $z/h \geq 1$, thereby promoting the increasing importance of $u^{\prime}$ fluctuations on the momentum transport as the heights approach the canopy top. 

Hitherto, the statistics of velocity fluctuations ($u^{\prime}$ and $w^{\prime}$) and momentum ($u^{\prime}w^{\prime}$) are presented for heights within and above the GoAmazon canopy. The growing importance of ejection and sweep motions with increasing heights and their differing contributions towards the time-fractions and momentum fluxes, lead us to ask, \emph{whether there is any particular time scale involved that differentiates the ejection and sweep characteristics observed so far}?. Answering such a question is not straightforward, since, during quadrant analysis any time dependence is masked and therefore no information can be obtained about the time scales of the associated turbulence structures. Additionally, the ejection and sweep events are not distributed uniformly over time but rather appear intermittently \citep{narasimha2007turbulent}. To overcome these issues, persistence analysis is proposed, which can evaluate the time scales of the intermittently-occurring momentum transport events.

\subsection{Persistence timescales}
\label{persistence}
As the logic dictates, we first employ persistence analysis to characterize the time scales of the streamwise and vertical velocity fluctuations ($u^{\prime}$ and $w^{\prime}$). Thereafter, to study how the two interacting signals $u^{\prime}$ and $w^{\prime}$ determine the time scales of momentum transporting events, a quadrant decomposition of persistence time scales is carried out. A detailed summary on the concepts of persistence analysis and its various implementation strategies are presented in \S\,\ref{methods}. In what follows, we discuss how the distributions of normalized persistence time scales behave across the nine heights within and above the canopy, when one considers the velocity and momentum flux signals.

\begin{figure}
\centering
\hspace*{-2.1in}
\includegraphics[width=1.5\textwidth]{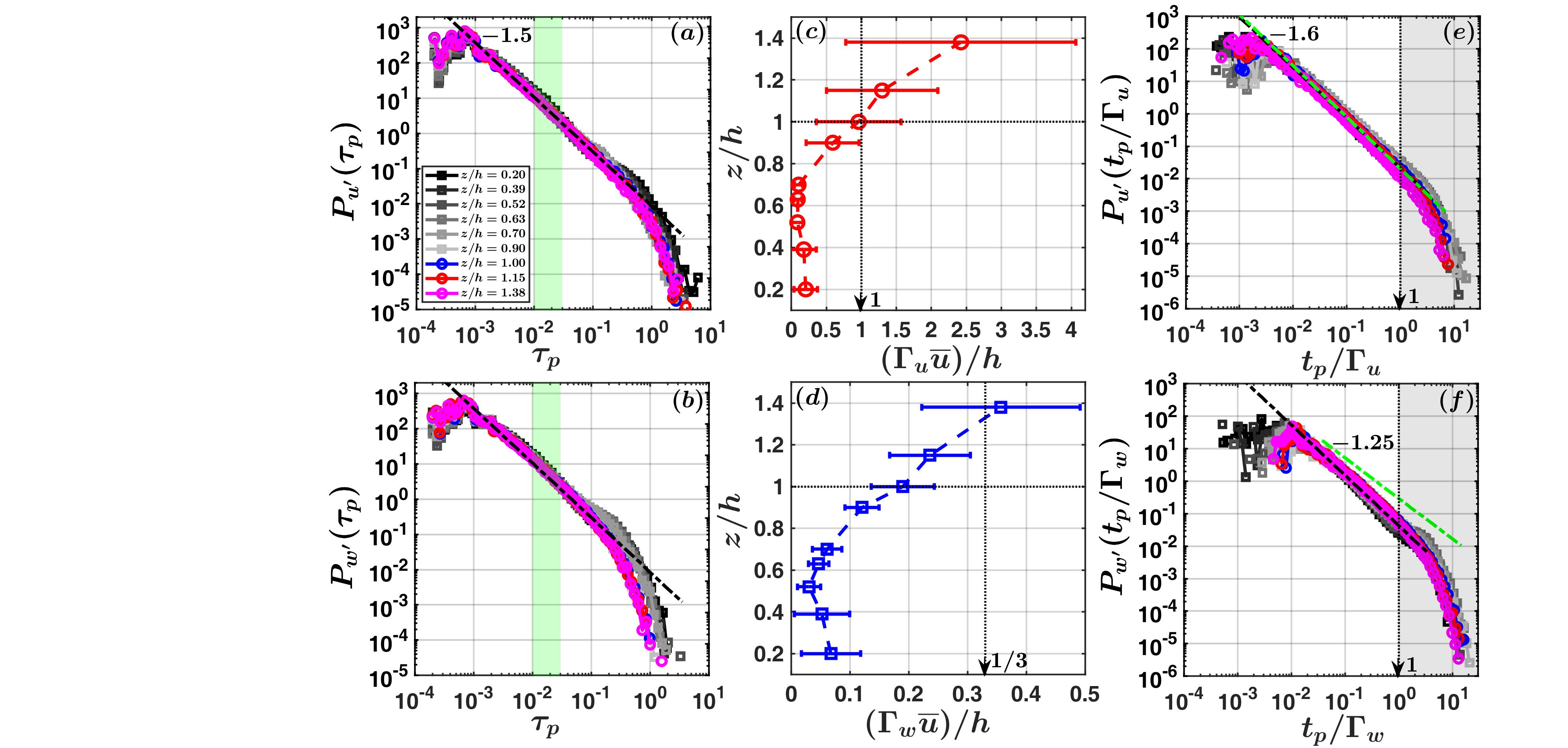}
 \caption{The persistence PDFs are shown for the events in (a) $u^{\prime}$ and (b) $w^{\prime}$, corresponding to the heights within and above the canopy. The normalized persistence time scales $(t_{p}u_{*})/h$ are denoted as $\tau_{p}$ and their PDFs are constructed over the ensemble of 93 half-hour blocks from near-neutral conditions. A power-law of exponent $-1.5$ is fitted to the persistence PDFs of $u^{\prime}$ events and shown as dash-dotted black lines. The green shaded regions in (a) and (b) denote the range of $\overline{\tau_{p}}$, over which it varies across the nine heights. In (c) and (d), the vertical profiles of normalized integral time scales ($\Gamma_{x}$, where $x=u$ or $w$) are shown for $u^{\prime}$ and $w^{\prime}$ signals, respectively. (e) and (f) depict the persistence PDFs of $u^{\prime}$ and $w^{\prime}$, but the time scales ($t_{p}$) are normalized by $\Gamma_{x}$. The gray shaded regions in (e) and (f) correspond to time scales $t_{p}/\Gamma_{x} \geq 1$. The power-laws with exponents $-1.6$ and $-1.25$ (dash-dotted green lines) are the fits redrawn from the SLTEST data \citep{chowdhuri2020persistence_a}.}
\label{fig:4}
\end{figure}

\subsubsection{Persistence PDFs of velocity fluctuations}
\label{pers_PDF}
In figure \ref{fig:4}a--b, the persistence PDFs of $u^{\prime}$ and $w^{\prime}$ signals are presented in log-log plots for all the nine measurement heights. Note that, before plotting the PDFs, the persistence time $t_{p}$ was scaled with a global time scale $h/u_{*}$ and denoted as $\tau_{p}$. Since $h/u_{*}$ is height-invariant, such a scaling of $t_{p}$ would allow one to investigate how $P(\tau_{p})$ changes with height. This scaling also has a distinct advantage if one compares the results with \citet{chamecki2013persistence}, who used a similar strategy to normalize the $u^{\prime}$ and $w^{\prime}$ persistence PDFs while studying non-Gaussian turbulence in canopy flows. To distinguish the heights within and above the canopy, an identical color scheme of figure \ref{fig:3} is used in figure \ref{fig:4} as well. 

By inspecting $P_{u^{\prime}}(\tau_{p})$ and $P_{w^{\prime}}(\tau_{p})$, one could immediately notice that these PDFs follow a power-law for nearly two decades up to a time scale $\tau_{p}<0.1$, and then there is an exponential drop. Although not explicitly mentioned, the plots of \citet{chamecki2013persistence} suggest the same limit for a power-law distribution in the persistence PDFs of $u^{\prime}$ and $w^{\prime}$. Moreover, \citet{chamecki2013persistence} pointed out that in his dataset -- at short time scales -- the persistence PDFs of either $u^{\prime}$ or $w^{\prime}$ did not display any substantially different behaviour with increasing heights. For the GoAmazon data, we also observe the same up to the time scales $\tau_{p}<0.1$.  

Interestingly, the inter-event distribution times between earthquakes have also been reported to follow a power-law distribution and the range is found to be limited up to the mean values of such times \citep[e.g.,][]{kumar2020interevent}. However, for canopy turbulence, the threshold timescale ($\tau_{p}=0.1$) up to which the power-law exists in $P_{u^{\prime}}(\tau_{p})$ and $P_{w^{\prime}}(\tau_{p})$ is nearly ten times larger than the mean values of $\tau_{p}$, denoted as $\overline{\tau_{p}}$. To illustrate this point, the green-shaded regions in figure \ref{fig:4}a--b indicate the range of possible values over which $\overline{\tau_{p}}$ would vary when all the nine measurement heights are considered together. As one could notice from the green-shaded regions, for both velocity signals, $\overline{\tau_{p}}$ displays little to no variations with height ($\overline{\tau_{p}}=$ 0.01 to 0.03) with the values being significantly smaller than $\tau_{p}=0.1$. This apparent difference with earthquakes suggests that the characteristics of the persistence PDFs are sensitive to the details of the physical system for which they are evaluated. Next, we provide further evidence that this sensitive dependence of persistence PDFs on the configurations of the underlying system remains true even when one considers a turbulent flow setting.  

For instance, the results of \citet{chamecki2013persistence} were obtained for a near-neutral turbulent flow over a corn canopy having $h=$ 2.05 m and an LAI of 3.3 m$^2$ m$^{-2}$. However, unlike \citet{chamecki2013persistence}, we do not observe any significant difference in the features of persistence PDFs when positive and negative fluctuations in $u^{\prime}$ and $w^{\prime}$ are considered separately. From figure S1a--b (electronic supplementary material) one could infer that the PDFs $P_{u^{\prime}>0}(\tau_{p})$ and $P_{u^{\prime}<0}(\tau_{p})$ behave almost similar to $P_{u^{\prime}}(\tau_{p})$ in figure \ref{fig:4}a. An identical conclusion can be drawn if one considers $w^{\prime}$ signals (figure S1c--d). Therefore, the statistical character of the persistence time scales of $u^{\prime}$ and $w^{\prime}$ signals appear to be different between the GoAmazon canopy and corn canopy of \citet{chamecki2013persistence}. Given this premise, one may additionally ask whether this difference also gets reflected in the power-law exponents of the persistence PDFs, an aspect that deserves a thorough investigation. 

In our case, since the power-law portions of persistence PDFs collapse with height, a single exponent can be determined by performing a linear regression for the range 10$^{-3}$ $\leq \tau_{p} \leq$ 10$^{-1}$ on the log-log plots of $P_{u^{\prime}}(\tau_{p})$. Although this is regarded as a standard practice \citep[e.g.,][]{katul1994conditional,chowdhuri2020persistence_a,Santiago2022}, estimating the accuracy of the power-law exponent is not a trivial task and involves a lot of complicated steps \citep{clauset2009power}. Nevertheless, to tackle this issue, we adopted a simple procedure. We first carried out the linear regressions for all the nine heights, thereby providing nine values of power-law exponents associated with each $z/h$. Note that, for all such fittings, the $R^{2}$ values were more than 0.98. In the next step, to estimate the error we computed the spread (standard deviation) around the mean over nine values of power-law exponents, eventually yielding a value of $-1.5 \pm 0.03$. For illustration purposes, the black dash-dotted line in figure \ref{fig:4}a represents the mean, which is $-1.5$. A similar process was repeated for $P_{w^{\prime}}(\tau_{p})$, and no discernible difference in the power-law exponent was observed when compared with $P_{u^{\prime}}(\tau_{p})$ (black dash-dotted lines in figure \ref{fig:4}a--b). 

Since \citet{chamecki2013persistence} did not provide the values of power-law exponents, we compare our findings with other studies conducted on canopy turbulence \citep{cava2009effects,lee2011intermittency,cava2012role} and on an aerodynamically smooth atmospheric surface layer (ASL) flow. Specifically, for the ASL flow, we directly use the Surface Layer Turbulence and Environmental Science Test (SLTEST) experimental dataset whose details are underlined in \citet{chowdhuri2020persistence_a,chowdhuri2019revisiting,chowdhuri2020persistence_b}. The SLTEST site was flat, horizontally homogeneous barring any obstacles, and having a smooth surface with an aerodynamic roughness length less than 5 mm \citep{chowdhuri2020persistence_a,chowdhuri2019revisiting,chowdhuri2020persistence_b}. Therefore, the SLTEST dataset provides a perfect platform to compare the persistence properties with the GoAmazon site which has roughness elements in the form of a tall canopy. As noted in previous studies, the near-neutral runs from SLTEST experiment were selected based on the criterion $-z/L<0.2$ \citep{chowdhuri2020persistence_a,chowdhuri2019revisiting,chowdhuri2020persistence_b}. 

Without employing any scaling, the studies by \citet{cava2009effects} and \citet{lee2011intermittency} discovered that in a near-neutral canopy flow setup ($h=$ 28, 14, and 8.5 m and LAI$=$ 9.6, 3, and 1 m$^2$ m$^{-2}$) the persistence PDFs of velocity components showed a log-normal distribution. On the other hand, the studies by \citet{cava2012role} and \citet{chowdhuri2020persistence_a} used the Eulerian integral time scales ($\Gamma_{x}$, $x=u$ or $w$) to normalize $t_{p}$, due to the following physical reasons. First, the use of $\Gamma_{x}$ allows one to judge whether the $t_{p}$ values are smaller or larger than the inertial subrange, since the scales smaller than $\Gamma_{x}$ are generally associated with such behaviors \citep{kaimal1994atmospheric,manshour2016interoccurrence}. Second, the power-law exponents of the persistence PDFs are typically explained in a framework of self-organized criticality and connected with the spectral slope in inertial subrange, subjected to corrections accounting for small-scale intermittency \citep{cava2012role,chowdhuri2020persistence_a,huang2021velocity}. Therefore, it is prudent to use the values of $\Gamma_{u}$ and $\Gamma_{w}$ for renormalizing the persistence time scales, $t_{p}$.

We compute the integral time scales by following the standard procedure that involves integrating the autocorrelation coefficients of any turbulent signal up to the lags until the first zero-crossing occurs \citep{li2012mean,huang2021velocity}. In appendix \ref{app_C}, the $u$ and $w$ spectral information are presented from where one can confirm that the time scales smaller or larger than the integral scales do indeed correspond to the inertial subrange or energy-production range (figure \ref{fig:A3}d--e). Furthermore, to check whether our estimates of $\Gamma_{u}$ and $\Gamma_{w}$ match with \citet{finnigan2000turbulence}, we convert those to normalized length scales using local mean wind speed ($\overline{u}$) and $h$. The vertical profiles of these quantities are shown in figure \ref{fig:4}c--d, where the error bars denote the spread around the ensemble mean. Typically, the profiles behave similar to \citet{finnigan2000turbulence} with the integral scales of $u^{\prime}$ being substantially larger than $w^{\prime}$. Moreover, at $z/h=1$, the integral length scale of $u^{\prime}$ remains equal to the canopy height (figure \ref{fig:4}c) whereas for $w^{\prime}$ a departure is observed from the stipulated value of $h/3$ (figure \ref{fig:4}d). We next describe the features of persistence PDFs when the timescales are scaled as $t_{p}/\Gamma_{x}$.

The overall character of $u^{\prime}$ and $w^{\prime}$ persistence PDFs under the new scaling scheme (figure \ref{fig:4}e--f) does not differ much from the old one (figure \ref{fig:4}a--b). The power-laws are followed up to the time scales $t_{p}<\Gamma_{x}$ and its exponents remain similar as before ($-1.5$). This is in sharp contrast to the momentum flux events, for which the effect of different scaling on its persistence time scales remain quite significant (discussed in \S\,\ref{alr_scale}). If the power-law exponent of $-1.5$, as obtained for the GoAmazon dataset, is compared with \citet{cava2012role} who considers many different canopy types, a large discrepancy could be noted. For instance, from \citet{cava2012role} it becomes evident that depending on the canopy type ($h=$ 16-28 m and LAI$=$ 7-9.6 m$^2$ m$^{-2}$) the exponents of $P_{u^{\prime}}(t_{p}/\Gamma_{u})$ and $P_{w^{\prime}}(t_{p}/\Gamma_{w})$ can vary between $-1.6$ to $-1.9$, with often the two not being the same. 

On the other hand, to assess the same in a near-neutral ASL flow over an aerodynamically smooth surface, we overlay $P_{u^{\prime}}(t_{p}/\Gamma_{u})$ and $P_{w^{\prime}}(t_{p}/\Gamma_{w})$ from the SLTEST experiment \citep{chowdhuri2020persistence_a} on the ones already obtained from GoAmazon canopy. This result is presented in figure S2a--b (electronic supplementary material), and also in figure \ref{fig:4}e--f where we only show the power-law fits from the SLTEST dataset as green dash-dotted lines. For the near-neutral SLTEST dataset, \citet{chowdhuri2020persistence_a} estimated the power-law exponents of $P_{u^{\prime}}(t_{p}/\Gamma_{u})$ and $P_{w^{\prime}}(t_{p}/\Gamma_{w})$ to be equal to $-1.6$ and $-1.25$, respectively. However, when those are overlaid on the GoAmazon dataset displaying an exponent of $-1.5$, there is hardly any difference for the $u^{\prime}$ signal (figure \ref{fig:4}e and figure S2a). But for the $w^{\prime}$ signal, a conspicuous difference is noted between the GoAmazon and SLTEST datasets (figure \ref{fig:4}f and figure S2b). 

From the above discussion, it is apparent that the power-law exponents of the persistence PDFs are far from universal owing to their dependence on the surface types over which a near-neutral atmospheric turbulent flow is studied. Other than the small time scales, the behaviour of the persistence PDFs at large time scales (energy-production scales) is also of interest. For instance, at time scales $\tau_{p} \geq 0.1$ or $t_{p} \geq \Gamma_{w}$, a bulge is noticed in the persistence PDFs of $w^{\prime}$, corresponding to the heights within the canopy (figure \ref{fig:4}b and f). However, for $u^{\prime}$, such a distinctive feature remains largely absent (figure \ref{fig:4}a and e). To examine the tails of the persistence PDFs in greater detail, one can investigate the complementary cumulative distribution functions (CCDF).

The CCDFs are denoted as $1-F_{x}(\mathcal{T})$, where $F_{x}(\mathcal{T})$ represents the conventional CDF ($x=u^{\prime},w^{\prime}$) while $\mathcal{T}$ indicates the normalized time scales $\tau_{p}$ or $t_{p}/\Gamma_{x}$. The results related to the CCDFs of $\tau_{p}$ and $t_{p}/\Gamma_{x}$ for both the velocity signals are shown in figure S3 (electronic supplementary material). At large time scales ($\tau_{p} \geq 0.1$ or $t_{p} \geq \Gamma_{u}$), the CCDFs of $u^{\prime}$ are found to be weakly dependent on heights (figure S3a and c). Conversely, for $w^{\prime}$, at $\tau_{p} \geq 0.1$, a clear separation is observed in their CCDFs between the heights within and above the canopy (figure S3b). It is interesting to note that this separation becomes a little less obvious when the $t_{p}$ values are scaled with $\Gamma_{w}$ (figure S3d). 

So far, the deliberations rendered on the $u^{\prime}$ and $w^{\prime}$ persistence characteristics develop an intriguing premise regarding the nature of $t_{p}$, related to four different quadrants of $u^{\prime}w^{\prime}$ signals. Certainly, one can ask,
\begin{enumerate}
    \item Due to the interaction between the persistence patterns in $u^{\prime}$ and $w^{\prime}$ signals, how $P(t_{p})$ would behave when the four different quadrants of the momentum-transporting events are considered separately?
    \item How the differing influence of heights on $P_{u^{\prime}}(t_{p})$ and $P_{w^{\prime}}(t_{p})$ affect the vertical evolution of the persistence patterns in $u^{\prime}w^{\prime}$ signals and whether any quadrant-wise change is observed in such patterns?
    \item Does there exist any particular scaling which can collapse the persistence PDFs of $u^{\prime}w^{\prime}$ signals, when all the nine measurement heights are considered together?
\end{enumerate}

\subsubsection{Quadrant decomposition of the persistence time scale}
\label{quad_decomp}
\begin{figure}
\centering
\hspace*{-1in}
\includegraphics[width=1.3\textwidth]{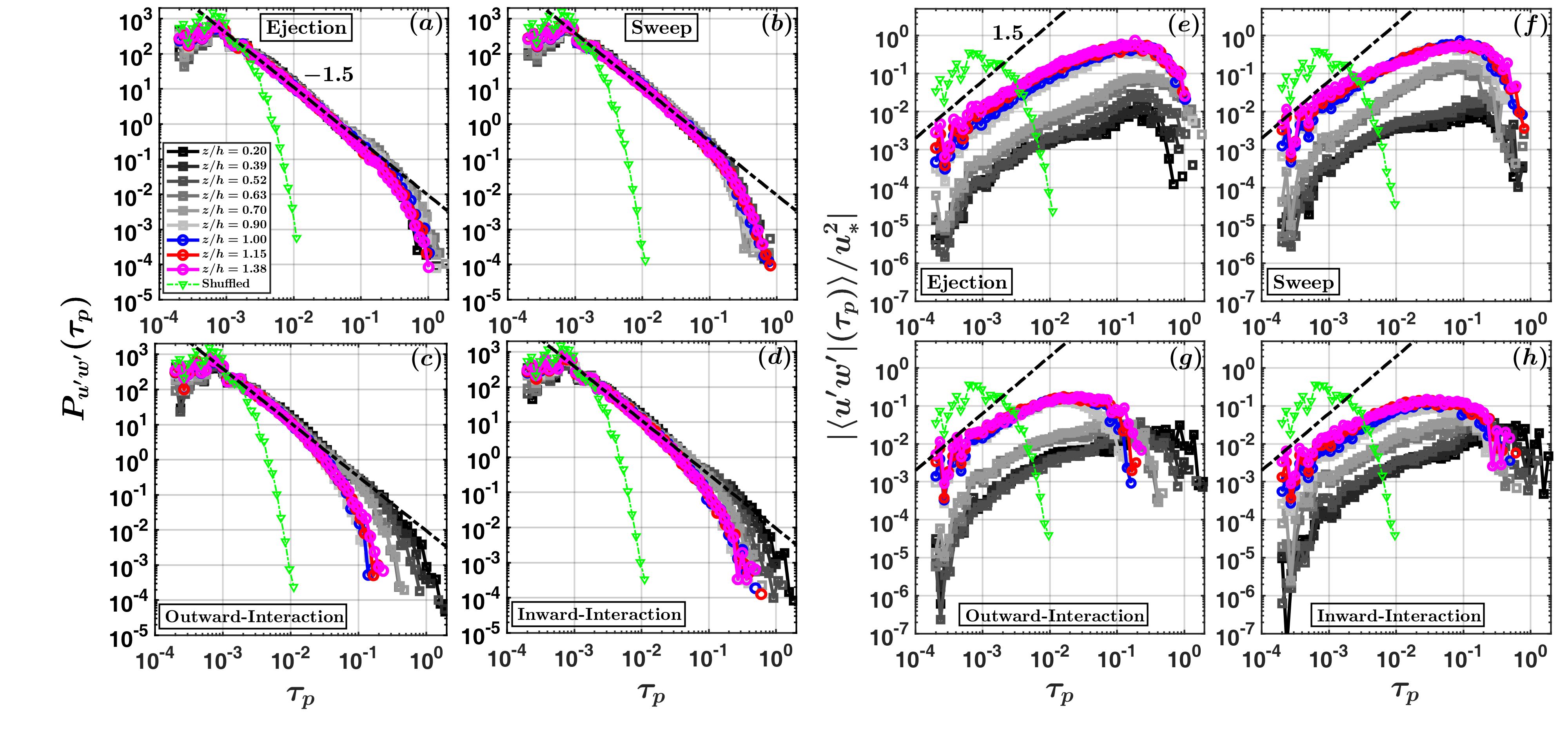}
 \caption{The persistence PDFs of $\tau_{p}$ are shown for the four different $u^{\prime}$-$w^{\prime}$ quadrant events, namely (a) ejection, (b)sweep, (c) outward-, and (d) inward-interaction, corresponding to the heights within and above the canopy. The dash-dotted green lines with triangle markers denote the results from a randomly-shuffled sequence of the $u^{\prime}$ and $w^{\prime}$ signals. In (e)--(h), the momentum flux distributions are plotted against $\tau_{p}$, depicted separately for the four different quadrants. Absolute values are used for the momentum flux distributions to clearly demonstrate the differences in their characteristics as the heights change.}
\label{fig:5}
\end{figure}

To explore these research questions, in figure \ref{fig:5}a--d we present the persistence PDFs of the momentum-transporting events, separately for the four different $u^{\prime}$-$w^{\prime}$ quadrants (table \ref{tab:1}). To compute these PDFs, the individual residence times in four quadrants are evaluated (see the graphical illustration in figure \ref{fig:1}c), and thereafter have been normalized with a particular time scale. Similar to $u^{\prime}$ and $w^{\prime}$ signals, we initially use the global time scale $h/u_{*}$ to normalize $t_{p}$ ($\tau_{p}$) so that the height variations can be clarified better. Note that, although the time scales of different quadrant events is investigated for the first time in canopy turbulence, there are a handful of studies in low-$Re$ engineering flows where such an analysis has been undertaken albeit in a different form. For instance, \citet{lozano2012three} employed a concept similar to persistence to study the sizes of the positive and negative momentum transport events in a direct numerical simulation (DNS) of a turbulent channel flow. In their work, the authors defined the sizes of the events as connected regions in a three-dimensional space.   

From figure \ref{fig:5}a--b, which depict $P_{u^{\prime}w^{\prime}}(\tau_{p})$ from ejection and sweep quadrants, one could notice that these PDFs remarkably collapse on to each other for all the nine heights within and above the canopy. The goodness of the collapse can even be verified from the CCDFs of the same, shown in figure S4a--b (electronic supplementary material). This finding is non-trivial due to the reasons underlined next. First and foremost, irrespective of the fact that at heights deep within the canopy almost no streamwise momentum is transported ($\overline{u^{\prime}w^{\prime}} \to$ 0, figure \ref{fig:2}c and e), the time scales of ejection and sweep events display a height-invariance. To confirm whether this outcome is a consequence of coherent structures in canopy flows, we generate randomly-shuffled surrogates of $u^{\prime}$ and $w^{\prime}$ signals and recompute $P_{u^{\prime}w^{\prime}}(\tau_{p})$. To briefly explain the shuffling procedure, one can select the time series of any signal and then operate a random permutation to disrupt the underlying temporal arrangement, thereby creating a surrogate dataset that does not possess any relationship among the signal data points \citep{chowdhuri2021visibility}. In this process, the signal's PDF remains precisely conserved although the appearance of the data points become random.  

We find that the persistence PDFs generated from randomly-shuffled sequences of $u^{\prime}$ and $w^{\prime}$ (green dash-dotted lines with triangle markers) markedly disagree with the observed ones (except for smallest $\tau_{p}$ values, which most likely represent noise) at all the nine heights. Since any temporal coherence is destroyed through random-shuffling, the disagreement of $P_{u^{\prime}w^{\prime}}(\tau_{p})$ with its randomly-shuffled surrogate physically suggests that the coherent structures, which constitute the ejection and sweep motions, extend their footprints deep within the canopy. Despite the presence of such coherent motions at heights deep inside the canopy, they do not transport any momentum as almost all are absorbed by the foliage \citep{dwyer1997turbulent}. In the words of \citet{townsend1961equilibrium} and \citet{bradshaw1967inactive} these motions could be referred to as the inactive ones since they have a negligible impact on the flux transport, i.e., contributing insignificantly to the total flux.

Moreover, up to $\tau_{p}<0.1$, the same power-law with an exponent $-1.5$, as observed in $P_{u^{\prime}}(\tau_{p})$ and $P_{w^{\prime}}(\tau_{p})$, also appears in the persistence PDFs of $u^{\prime}w^{\prime}$ events from ejection and sweep quadrants (figure \ref{fig:5}a--b). This power-law exponent does not match with the observations of \citet{katul1994conditional} and \citet{chowdhuri2020persistence_b} from a near-neutral ASL flow. Both of these studies found, at short time scales, the persistence PDFs of momentum- and heat-transporting events follow a power-law with an exponent $-1.4$. For a visual demonstration, one can consult figure S2c (electronic supplementary material) keeping in mind that the scaling of $t_{p}$ is with $\Gamma_{w}$. In that plot, the persistence PDFs of $u^{\prime}w^{\prime}$ events (considering all the four quadrants together) from the SLTEST dataset are overlaid on the GoAmazon one and a clear difference between the two could be noted in the power-law section. 

Apart from these high-$Re$ atmospheric flows ($Re \approx 10^7$), \citet{kailasnath1993zero} found that in the inertial layer of a moderate $Re$ (between 2800 to 13000) flat-plate boundary layer, the persistence PDFs of $u^{\prime}w^{\prime}$ signal closely followed the same PDFs of $w^{\prime}$, with both displaying a nearly identical single exponential function. Therefore, no indication of a power-law was evident in their studies. Conversely, from the DNS results of a turbulent channel flow ($Re \approx 5000$), \citet{lozano2012three} found that the PDFs of streamwise lengths of negative momentum events (comprising of ejections and sweeps) followed a power-law distribution with an exponent between $-4/3$ and $-5/3$. These results together with the ones from ASL and canopy flows extend our previous conclusion on $u^{\prime}$ and $w^{\prime}$ persistence PDFs to the $u^{\prime}w^{\prime}$ signals. In other words, they reinforce the fact that the statistical character of these PDFs are indeed dependent on the flow type that is being studied. Physically, the reason behind such dependency may lie in the structural discrepancies between the roughness sublayer (RSL) of canopy flows and inertial sublayer (ISL) of a rough- or smooth-wall boundary layer. In the ISL of wall-bounded turbulent flows, the ejection and sweep motions are typically associated with hairpin structures \citep{adrian2000vortex,ganapathisubramani2003characteristics,hommema2003packet,adrian2007hairpin,fiscaletti2018spatial} and by employing persistence analysis one expects to provide insights into their time scales \citep{kailasnath1993zero}. However, the presence of a canopy in the RSL alters the structures of these hairpins \citep{finnigan2009turbulence}. Notwithstanding different opinions on how exactly such modification occurs \citep{bailey2016creation}, it appears that these modified hairpin structures cause a difference in the persistence results when one compares the RSL and ISL flows.    

Continuing with figure \ref{fig:5}, one can see that contrary to the ejection and sweep quadrants, $P_{u^{\prime}w^{\prime}}(\tau_{p})$ from the counter-gradient quadrants (outward- and inward-interaction) show a considerable amount of height-dependency (figure \ref{fig:5}c--d). Such a sensitive dependence on $z$ is more clearly visible in the corresponding CCDFs (figure S4c--d, electronic supplementary material). By further comparing the features of persistence PDFs among four different quadrants, we find that at heights deep within the canopy, the occurrences of large-duration events from counter-gradient quadrants are statistically as likely as the ones from the ejection and sweep quadrants. On the other hand, at $z/h \geq 1$, $P_{u^{\prime}w^{\prime}}(\tau_{p})$ of outward- and inward-interaction quadrants drop off quite rapidly at large $\tau_{p}$ values as compared to the co-gradient quadrants. This observation suggests, at heights above the canopy, the long-duration events are primarily dominated by the ejection and sweep motions. Note that such separation in persistence PDFs between the co- and counter-gradient quadrants mainly happens at large $\tau_{p}$ values ($\tau_{p}>0.1$), while at small $\tau_{p}$ ($\tau_{p}<0.1$) the quadrant-wise distinction remains hardly noticeable. Unlike the exponents, this finding is in qualitative agreement with the results obtained from near-neutral ASL flows \citep{chowdhuri2019revisiting,chowdhuri2020persistence_b} and with the DNS of a turbulent channel flow \citep{lozano2012three}.

Overall, the quadrant-decomposed features of persistence help one to comprehend the interaction between the structures in $u^{\prime}$ and $w^{\prime}$ signals, generating the dynamics of the momentum-transporting events. While plotting the persistence PDFs corresponding to the ejection and sweep quadrants, we note these PDFs are height-invariant. However, the same is not true for the outward- and inward-interaction quadrants, for which a strong sensitivity to the heights are noticed. Concerning the velocity signals, the PDFs of persistence time scales associated with $u^{\prime}$ show a little to no height dependence (figure \ref{fig:4}a). On the other hand, a clear bulge is observed in the persistence PDFs of $w^{\prime}$ at time scales $\tau_{p} \geq 1$ (figure \ref{fig:4}b). Therefore, at those time scales, the persistence PDFs of $w^{\prime}$ differ between the in- and above-canopy measurement locations, meaning they depend on $z$ unlike the $u^{\prime}$ one. One may ask why such bulge observed in the persistence PDFs of $w^{\prime}$ signal does not get reflected in the time scale distributions of co-gradient motions? Given height-invariance is the common factor, we hypothesize that the distributions of the persistence time scales of ejection and sweep motions are most likely dominated by the positive and negative patterns in the $u^{\prime}$ signal. In contrast, the patterns in the $w^{\prime}$ signal exert their influences to decide the nature of the persistence PDFs of counter-gradient quadrants, ultimately resulting in the observed differences with heights. 

Notwithstanding these valuable insights, one needs to remember that the persistence analysis provides information on the time scale distribution of the flux events but not on the flux values themselves. This limitation arises, because the amplitude variations are not accounted for in the persistence PDFs. Hence, to establish the role of quadrant-wise persistence patterns towards the momentum transport, additional analysis is required (see \S\,\ref{flux_contributions}). 

\subsubsection{Flux contributions from persistence patterns}
\label{flux_contr}
Figure \ref{fig:5}e--h individually depict the quadrant contributions to the momentum flux against the normalized persistence time scale $\tau_{p}$, corresponding to all the nine measurement heights. Similar to persistence PDFs, the momentum flux distributions are far apart from a random configuration, thus accounting for the dynamical flow structures. Moreover, upon considering amplitudes one can see that, notwithstanding the excellent collapse observed in the PDFs of ejection and sweep time scales (figure \ref{fig:5}a--b), momentum flux distributions against $\tau_{p}$ show a clear dependence on $z$ (figure \ref{fig:5}e--f). To emphasize such dependencies, absolute values of $\langle u^{\prime}w^{\prime}\vert(\tau_{p})\rangle/u^{2}_{*}$ are plotted in figure \ref{fig:5}e--h on a logarithmic scale.  

\begin{figure}
\centering
\hspace*{-1.55in}
\includegraphics[width=1.5\textwidth]{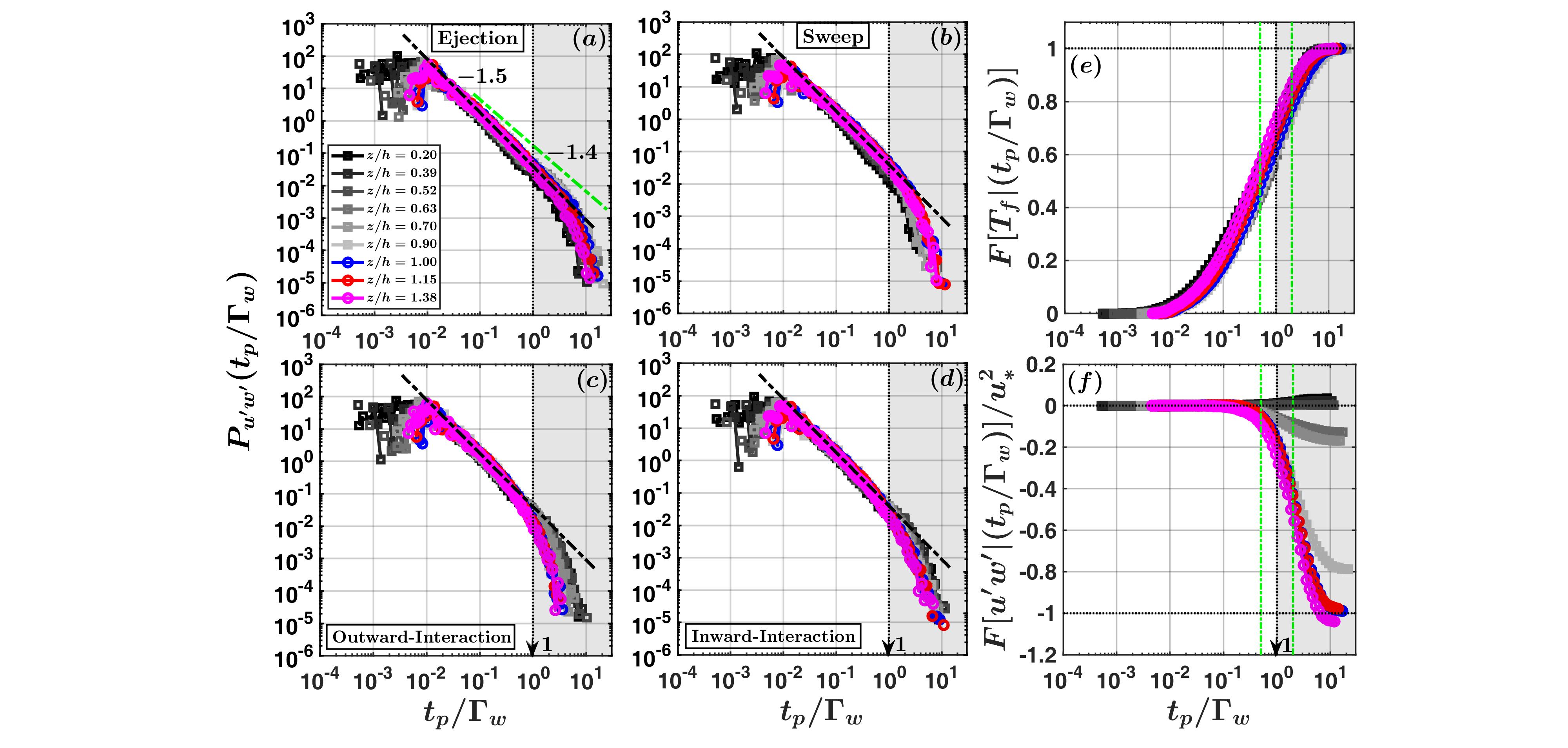}
 \caption{In (a)--(d), the persistence PDFs are shown for the four different $u^{\prime}$-$w^{\prime}$ quadrant events, but the time scales are scaled as $t_{p}/\Gamma_{w}$. In (a), power-law with exponent $-1.4$ (dash-dotted green lines) is shown, which is a fitting result to the SLTEST data as obtained from \citet{chowdhuri2020persistence_b}. In (e) and (f), cumulative contributions to the time fractions and momentum fluxes are shown against the time scales ($t_{p}/\Gamma_{w}$) of $u^{\prime}w^{\prime}$ events. The two green coloured dash-dotted vertical lines on (e) and (f) denote the time scales $t_{p}/\Gamma_{w}=$ 0.5 and 2, respectively. The gray shaded regions correspond to time scales $t_{p}/\Gamma_{w} \geq 1$.}
\label{fig:6}
\end{figure}

At time scales $\tau_{p}<0.1$, flux distributions corresponding to ejections and sweeps, display a distinct power-law behaviour (figure \ref{fig:5}a--b). However, in these plots, the power-law exponents appear to be different from $1.5$ (shown in dash-dotted black lines). There are a couple of consequences of this result --- the first one is from a modelling standpoint while the other one is speculative. Regarding the first, persistence PDFs and associated flux distributions display exponents whose absolute values do not match. As discussed in \S\,\ref{flux_contributions}, this indicates that the information about persistence is not enough and amplitude variability of the instantaneous momentum fluxes should be taken into account while computing the flux contributions from individual events of different time scales. We shed further light on this issue in \S\,\ref{scale_sep}. 

Considering the second point, which is related to Brownian motions \citep{castellanos2013intermittency}, one may argue that the presence of amplitudes causes a departure from a local gradient-diffusion parameterization \citep{banerjee2017connecting,li2022modeling} and thus precluding the phenomenological use of a turbulent diffusivity (analogous to Brownian motion) to describe momentum transport. However, more evidences are needed to support this argument and providing those remain beyond the scope of this study. Interestingly, unlike ejections, for the sweep motions, the slopes of $\langle u^{\prime}w^{\prime}\vert(\tau_{p})\rangle/u^{2}_{*}$ gradually become steeper as $z/h$ approaches 1. This behaviour of the sweep events documents a change in the small-scale dynamics from heights within the canopy to above it. 

At larger time scales, the power-law behaviour disappears with $\langle u^{\prime}w^{\prime}\vert(\tau_{p})\rangle/u^{2}_{*}$ attaining peaks at $\tau_{p} \approx 0.2$ and $\tau_{p} \approx 0.1$ for the ejection and sweep quadrants, respectively (figure \ref{fig:5}e--f). On the other hand, no clear peaks are noticed for the counter-gradient quadrants and the flux distributions remain mostly confined to the scales $\tau_{p}<0.1$ at $z/h \geq 1$ (figure \ref{fig:5}g--h). This is consistent with figure \ref{fig:5}c--d where it was shown that at heights above the canopy (contrary to the ones within), the large time scale ($\tau_{p}>0.1$) events become highly infrequent for outward- and inward-interaction quadrants. As a whole, the features of figure \ref{fig:5} make one wonder whether the apparent distinction that exists between the persistence properties of co-gradient and counter-gradient quadrants could be resolved if a different scaling is used other than $h/u_{*}$. Put differently, it remains unclear if the persistence time scales of co-gradient motions scale differently than the counter-gradient ones. 

\subsubsection{An alternate scaling with integral scales}
\label{alr_scale}
Since the persistence time scales of $u^{\prime}w^{\prime}$ events are a result of the interaction between the positive and negative patterns in the component signals ($u^{\prime}$, $w^{\prime}$), there could be two alternate possibilities to scale the $t_{p}$ of four different quadrants. From our foregoing discussion in \S\,\ref{quad_decomp}, it appears that the time scales associated with the co- or counter-gradient motions may scale separately with $\Gamma_{u}$ or $\Gamma_{w}$.

To confirm this hypothesis, one first attempts to scale $t_{p}$ with $\Gamma_{u}$. However, under such scaling, both the persistence PDFs and CCDFs (figure S5 in the electronic supplementary material) exhibit a poorer collapse with heights than what was being reported in figure \ref{fig:5}a--d. For instance, regarding the counter-gradient quadrants, one notices that the persistence time scales inside the canopy well exceed $\Gamma_{u}$, while the opposite remain true for the heights $z/h \geq 1$ (figure S5 in the electronic supplementary material). On the other hand, the distributions of persistence time scales collapse the best for all the four quadrant events when normalized by the integral time scale of the vertical velocity ($\Gamma_{w}$). For demonstration purposes, the persistence PDFs of $t_{p}/\Gamma_{w}$ are presented in figure \ref{fig:6}a--d while the corresponding CCDFs can be found in figure S6 (electronic supplementary material). To denote the apparent mismatch with the SLTEST dataset, the dash-dotted lines (black and green) in figure \ref{fig:6}a indicate both the power-law exponents $-1.5$ and $-1.4$. 

The collapse of quadrant-wise persistence PDFs under the $\Gamma_{w}$-scaling physically suggests that, the eddy structures whose time scales are comparable to $\Gamma_{w}$, participate in the momentum transport (either actively or inactively) both within and above the GoAmazon canopy. In fact, one can argue that this result from persistence analysis adds significant value to the information obtained from the $u$-$w$ cospectra (see the discussion in \S\,\ref{event_sep}). Although previous researchers have commented that the integral scale of vertical velocity is comparable to the canopy height $h$ \citep{finnigan2009turbulence}, our results show if $h$ is used in place of $\Gamma_{w}$ the persistence PDFs of counter-gradient events do not collapse with heights. Contrarily, for ejections and sweeps, this difference is insignificant. While converting $\Gamma_{w}$ to a length scale Taylor's hypothesis is needed, which often is questionable in canopy turbulence \citep{finnigan2009turbulence,everard2021sweeping}. It could be possible that the discrepancy between the two scalings for the counter-gradient quadrants is a result of the failure of Taylor's hypothesis. Interestingly, the co-gradient quadrant results remain unaffected by this. Moreover, when normalized by $\Gamma_{w}$, the distinction between the power-law and exponential drop in $P_{u^{\prime}w^{\prime}}(t_{p}/\Gamma_{w})$ becomes quite evident at $t_{p}/\Gamma_{w}=1$, irrespective of the quadrant type.

From these newly-scaled persistence PDFs, we can conclude that the $u^{\prime}w^{\prime}$ events remain statistically self-similar up to the time scales $t_{p}/\Gamma_{w}<1$. However, such self-similarity gets destroyed at scales $t_{p}/\Gamma_{w} \geq 1$ due to the presence of large-scale coherent structures. Therefore, it can be stipulated that $t_{p}/\Gamma_{w}=1$ represents a threshold to separate two distinct event classes, one whose statistical characteristics show a power-law behaviour and one for which a deviation is noted from the power-law. Furthermore, to assess whether the statistics of momentum flux events remain sensitive to the choice of the threshold while separating the two event classes, one may consult figure \ref{fig:6}e--f. 

In figure \ref{fig:6}e--f, the cumulative contributions to the time-fractions ($F[T_{f} \vert (t_{p}/\Gamma_{w})]$) and momentum fluxes ($F[u^{\prime}w^{\prime} \vert (t_{p}/\Gamma_{w})]$) are shown, if one considers the persistence time scales of all the four quadrants together. In order to explain the features better, we, as an example, focus on the $x$-axis value $t_{p}/\Gamma_{w}=1$. Corresponding to this time scale, the $y$-axis value in figure \ref{fig:6}e indicates how much fraction of a time would be occupied if only those events are taken into account whose time scales lie in the range $t_{p}/\Gamma_{w} \leq 1$. A similar inference could be drawn by focussing on figure \ref{fig:6}f but for the momentum flux contributions. Because the cumulative effects are presented, $y$-axes values in figure \ref{fig:6}e--f converge towards 1 and time-averaged momentum flux ($\overline{u^{\prime}w^{\prime}}/u^2_{*}$, figure \ref{fig:2}c), respectively. 

Regarding the sensitivity to the threshold, two vertical green dash-dotted lines in figure \ref{fig:6}e--f denote the time scales $t_{p}/\Gamma_{w}=0.5$ and $t_{p}/\Gamma_{w}=2$. By reading the values on the vertical axes associated with these scales, it can be identified how much the time fractions and flux contributions would have changed if instead of fixing the threshold at $t_{p}/\Gamma_{w}=1$, one had fixed it at 0.5 or 2. Upon completion of such task, nearly a $\pm 20\%$ change can be noted for both these quantities. Keeping this in mind, we next present results to address the connection between the persistence time scales and bulk statistical features of momentum transport reported in figures \ref{fig:2}--\ref{fig:3}. 

\subsection{Scale separation and amplitude variations}
\label{scale_sep}
\subsubsection{Short- and long-lived events}
\label{event_sep}
\begin{figure}
\centering
\hspace*{-1.8in}
\includegraphics[width=1.5\textwidth]{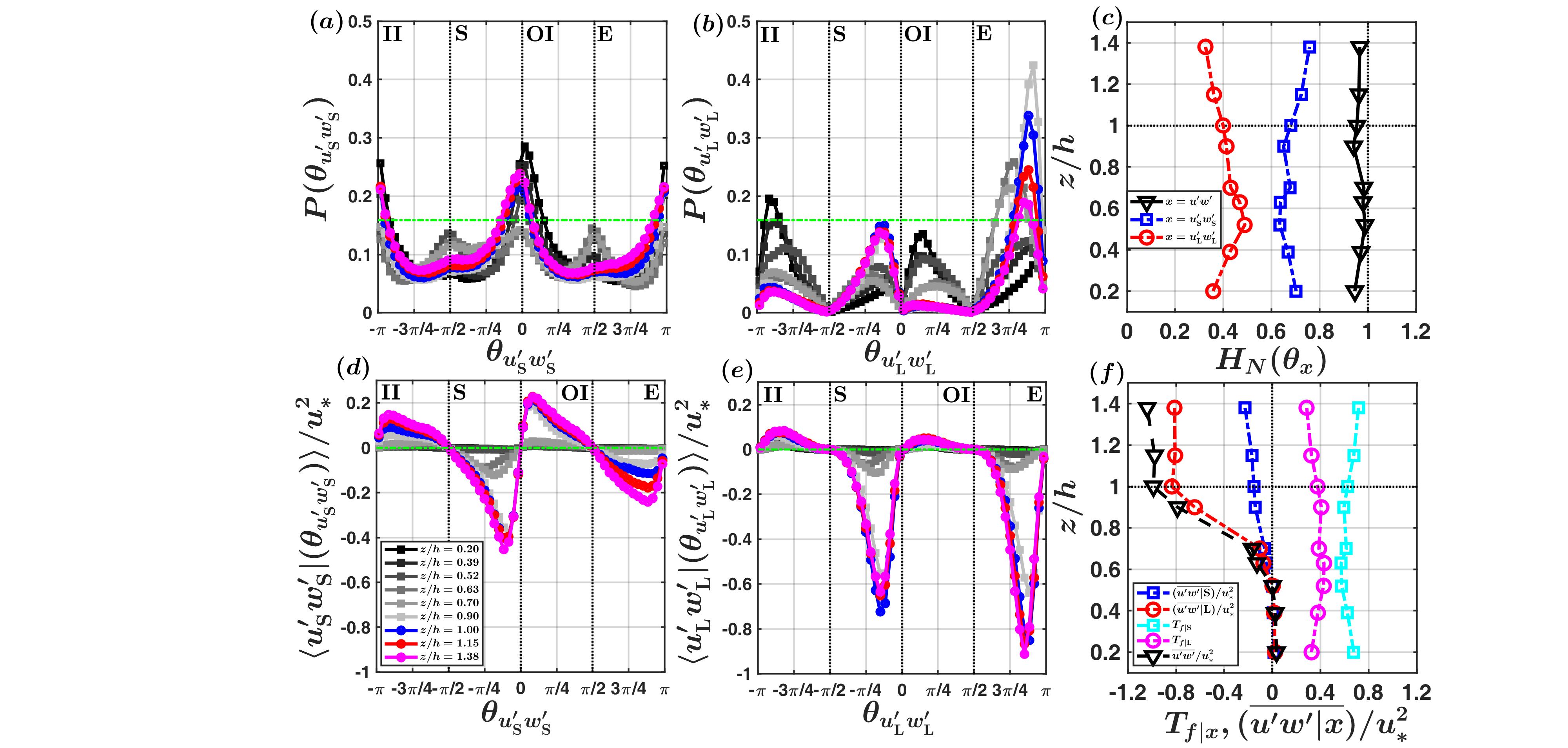}
 \caption{PDFs of quadrant angles are presented for events whose duration remains in the range (a) $t_{p}/\Gamma_{w}<1$ ($u^{\prime}_{\rm S}w^{\prime}_{\rm S}$) and (b) $t_{p}/\Gamma_{w} \geq 1$ ($u^{\prime}_{\rm L}w^{\prime}_{\rm L}$). In (c), normalized Shannon entropies of the quadrant angles are shown for the two event classes. In (d) and (e), quadrant distributions of the momentum flux are shown for the two event classes. The contributions to the time fractions and momentum fluxes from the short- (S) and long-lived (L) events are depicted in (f). }
\label{fig:7}
\end{figure}

In figure \ref{fig:7}, we demonstrate how the two event classes, distinguished by their normalized time scales $t_{p}/\Gamma_{w}<1$ (short-lived events) and $t_{p}/\Gamma_{w} \geq 1$ (long-lived events), affect the bulk properties of momentum transport (shown in figure \ref{fig:2}e--f and \ref{fig:3}f--g). To construct these plots, we conditionally sample the $u^{\prime}$ and $w^{\prime}$ values, pertaining to the events which persist for times smaller or larger than the threshold, $t_{p}/\Gamma_{w}=1$. Henceforth, we denote the signals sampled from the short- and long-lived events as $\{u^{\prime}_{\rm S}, w^{\prime}_{\rm S}\}$ and $\{u^{\prime}_{\rm L}, w^{\prime}_{\rm L}\}$, respectively. In this regard, one may note that, $u^{\prime}=u^{\prime}_{\rm S}+u^{\prime}_{\rm L}$, $w^{\prime}=w^{\prime}_{\rm S}+w^{\prime}_{\rm L}$, and $u^{\prime}w^{\prime}=u^{\prime}_{\rm S}w^{\prime}_{\rm S}+u^{\prime}_{\rm L}w^{\prime}_{\rm L}$. 

To investigate the impact of these two event classes on quadrant statistics, figure \ref{fig:7}a--b represent the phase angle PDFs, separately for the short- and long-lived events. These plots are designed in such a way that when summed over $P(\theta_{u^{\prime}_{\rm S}w^{\prime}_{\rm S}})$ and $P(\theta_{u^{\prime}_{\rm L}w^{\prime}_{\rm L}})$, one would recover the phase angle PDFs of figure \ref{fig:3}f. Similar to figure \ref{fig:3}f, we use 60 bins of $\theta_{u^{\prime}w^{\prime}}$ to compute $P(\theta_{u^{\prime}_{\rm S}w^{\prime}_{\rm S}})$ and $P(\theta_{u^{\prime}_{\rm L}w^{\prime}_{\rm L}})$. Irrespective of the event type, when 93 half-hour blocks are considered together, the samples corresponding to each quadrant angle PDFs turn out to be close to $10^5$, a significantly large number to ensure statistical robustness. 

As opposed to $P(\theta_{u^{\prime}_{\rm L}w^{\prime}_{\rm L}})$, one cannot recognize any apparent asymmetry among the four different quadrants in $P(\theta_{u^{\prime}_{\rm S}w^{\prime}_{\rm S}})$. Particularly, for the large-scale events, $P(\theta_{u^{\prime}_{\rm L}w^{\prime}_{\rm L}})$ values at heights all but the lowest three remain clearly skewed towards the ejection quadrants with respect to sweeps (figure \ref{fig:7}b). These PDF features can be further connected to the turbulence organization. For instance, an equipartition of the phase angle PDFs among the four different quadrants indicates that the organizational structure of turbulence associated with such events is close to random. In order to quantify the organizational state of these two event classes, we compute the normalized Shannon entropy of $\theta_{u^{\prime}_{\rm S}w^{\prime}_{\rm S}}$ and $\theta_{u^{\prime}_{\rm L}w^{\prime}_{\rm L}}$. Using information theory \citep{shannon1948mathematical,banerjee2018coherent,ghannam2020inverse}, the normalized Shannon entropy ($H_{N}$) of $\theta_{u^{\prime}w^{\prime}}$ can be defined as,
\begin{equation}
H_{N}(\theta_{u^{\prime}w^{\prime}})=-\frac{1}{\ln{(N_{\rm b})}}\sum_{i=1}^{N_{\rm b}} P_{i}(\theta_{u^{\prime}w^{\prime}}) \ln{[P_{i}(\theta_{u^{\prime}w^{\prime}})]},
\label{shannon}
\end{equation}
where $N_{\rm b}$ is the number of bins over which the $\theta_{u^{\prime}w^{\prime}}$ values are divided (60 in our case), and $P_{i}(\theta_{u^{\prime}w^{\prime}})$ is the probability of occurrence of a particular binned value $\theta_{u^{\prime}w^{\prime}}$. For a uniform distribution, $H_{N}(\theta_{u^{\prime}w^{\prime}})$ is equal to 1, and therefore the departure from unity could be regarded as a metric quantifying discrepancy with an equipartition configuration of $\theta_{u^{\prime}w^{\prime}}$. 

In figure \ref{fig:7}c, we plot the vertical profiles of $H_{N}$ values corresponding to the short- (blue lines with squares) and long-lived (red lines with circles) events. These profiles confirm that, relative to the short-lived events, the long-lived ones gradually become more organized as $z/h$ approaches 1. We also compare these values with $H_{N}(\theta_{u^{\prime}w^{\prime}})$ (black lines with triangles), obtained from the full signal without any scale separation. Since $H_{N}(\theta_{u^{\prime}w^{\prime}}) \approx 1$, one can conclude that, without separating the events based on their persistence time scales, it is impossible to extract information about the organization pattern of the coherent structures. After establishing this important point, to further unravel the event characteristics, we scrutinize the plots involving quadrant flux distributions in the phase angle space (figure \ref{fig:7}d--e). 

Similar to figure \ref{fig:7}a--b, the summed flux distributions of short- and long-lived events in figure \ref{fig:7}d--e yield the total, as shown in figure \ref{fig:3}g. Indeed, the flux distributions in the phase angle space portray a sharp contrast between the short- and long-lived events. Excluding the lowest three heights, sweep quadrants control the momentum transport at scales $t_{p}/\Gamma_{w}<1$ (figure \ref{fig:7}d). In fact, for the mid canopy heights ($z/h$ between 0.63 to 0.9), flux contributions from the short-lived sweep events greatly exceed the ejections. Accordingly, this finding suggests that the influence of sweep motions over the flux fractions at mid canopy heights (figure \ref{fig:2}e) is caused by those events, which persist for times $t_{p}/\Gamma_{w}<1$. On the contrary, the dominance of ejections in the time-fractions (figure \ref{fig:2}f) can be attributed to those events whose persistence time scales are $t_{p}/\Gamma_{w} \geq 1$. 

However, for $z/h$ values between 0.9 to 1.38, flux contributions from ejections and sweeps at scales $t_{p}/\Gamma_{w}<1$ remain significantly small when compared to its long-lived counterparts. Conversely, for the counter-gradient quadrants, in agreement with figure \ref{fig:5}g--h, they only exert their influence on momentum transport for the short-lived events while appearing almost absent at scales $t_{p}/\Gamma_{w} \geq 1$. To quantify these features, we investigate the contributions from these two event classes (considering all the four quadrants together) towards the time fractions and momentum fluxes with increasing $z/h$ (figure \ref{fig:7}f).

In figure \ref{fig:7}f, the quantity $T_{f \vert x}$ denotes the contributions to the time fractions from the events whose time scales are selected based on the condition $x$, i.e., whether $t_{p}<\Gamma_{w}$ (short-lived) or $t_{p}\geq \Gamma_{w}$ (long-lived). Similarly, $(\overline{u^{\prime}w^{\prime} \vert x})/u^{2}_{*}$ in figure \ref{fig:7}f indicates normalized contributions to the time-averaged momentum flux from the short- or long-lived events. We use different coloured lines to differentiate between the vertical profiles of time fractions and flux contributions in figure \ref{fig:7}f (see the legend). Since time fractions are positive definite quantities while the momentum fluxes are predominantly negative, the vertical zero line in figure \ref{fig:7}f separates the two. Furthermore, if the flux contributions are summed from these two event classes, one would obtain the total time-averaged momentum flux. To illustrate this point, we show black colored triangles in figure \ref{fig:7}f, which represent the vertical profile of $(\overline{u^{\prime}w^{\prime}})/u^{2}_{*}$ in figure \ref{fig:2}c without any error bars. 

From figure \ref{fig:7}f, one can see that the long-lived events occupy around 40\% percent of time, while the rest is dominated by the short-lived ones. This allows one to infer that not much change exists in the time fractions between these two event classes as both remain close to 0.5. However, flux contributions from these two events show remarkable differences. For instance, at heights with $z/h$ values between 0.9 to 1.38, almost all the momentum fluxes are carried within the long-lived events ($\approx$ 80\%), while the short-lived ones only contribute marginally ($\approx$ 20\%). The appearance of long-lived events as the primary carrier of momentum flux agrees well with the results obtained from near-neutral ASL flows \citep{chowdhuri2020persistence_b} and low-$Re$ DNS of turbulent channel flows \citep{lozano2012three,lozano2014time}. Inspite of not being conceptually the same, $u$-$w$ cospectra in figure \ref{fig:A3}f (appendix \ref{app_C}) indicates nearly all the momentum transport is associated with the periods larger than $\Gamma_{w}$. An analogous connection between the event-based and Fourier analyses was observed by \citet{lozano2012three} too for a turbulent channel flow. They mentioned that the length scales obtained from their event-based analysis could indeed be related to the spectrum of wall shear stress, thereby implying that such scales were the carriers of the Reynolds-stress fluctuations. Moreover, in the context of atmospheric turbulence, \citet{huang2021velocity} presented an analogy between the turbulence spectra and the shape of persistence PDFs. Therefore, from this perspective, the persistence time scales smaller than $\Gamma_{w}$ hardly contributing to the momentum is in qualitative agreement with the assumption of quasi-isotropic conditions in inertial-subrange turbulence \citep{Wyn72}.  

In general, long-lived events are the signatures of the coherent structures and particularly for the ISL of a wall-bounded flow, they have been linked to the presence of attached eddies or hairpin structures \citep{lozano2012three,lozano2014time}. Notwithstanding the fact that in the RSL of a canopy flow the hairpin structures get modified \citep{finnigan2009turbulence}, we do find an agreement between ISL and RSL flows when the contributions of long events to the momentum fluxes are considered. Since $u^{\prime}w^{\prime}$ persistence PDFs differ between the two flows but the flux contributions qualitatively match, this points towards a complicated role of amplitude variations in momentum transport. In fact, the amplitudes of the long-lived events should be substantially stronger than the short-lived ones so that, despite their similar contributions towards $T_{f}$, they nearly carry almost all the flux (figure \ref{fig:7}f). Therefore, from both figure \ref{fig:5} and \ref{fig:7}, the importance of amplitudes is abundantly evident. This outcome also raises a couple of questions whose answers are not trivial, such as:
\begin{enumerate}
    \item Does there exist a range of persistence time scales for which the effect of $u^{\prime}w^{\prime}$ amplitudes are felt most strongly while evaluating the event contributions to the momentum transport?
    \item Does the scale-wise amplitude variations in $u^{\prime}w^{\prime}$ depend on the quadrant type of the flux events? For example, is the amplitude variability associated with the time scales of ejection events appear stronger than the sweeps?
\end{enumerate}
To tackle these issues, we separate the effect of amplitude variations from persistence by developing a method that involves surrogate datasets.

\subsubsection{Amplitude variations}
\label{amp_var}
\begin{figure}
\centering
\hspace*{-1.45in}
\includegraphics[width=1.5\textwidth]{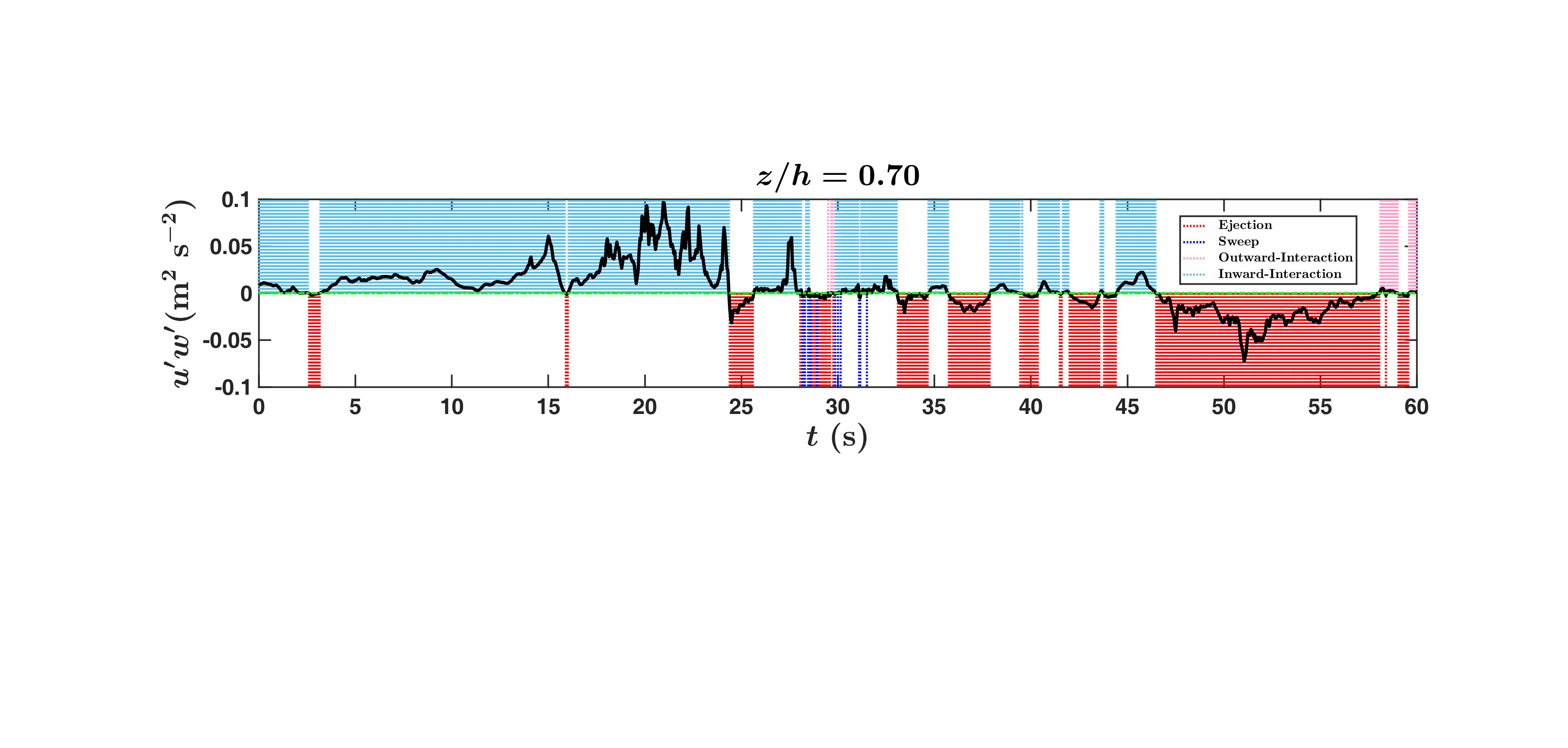}
\vspace{-4cm}
 \caption{A 60-s section time series of absolute values of $u^{\prime}w^{\prime}$ is shown to illustrate the concept of ignoring the amplitude variations in the flux values. The four different quadrant states are color-coded, as indicated in the legend.}
\label{fig:8}
\end{figure}

To explain our method, let us consider a 60-s time-series of the instantaneous momentum flux (shown as black lines in figure \ref{fig:8}) at a particular height, $z/h=0.7$. In figure \ref{fig:8}, the different quadrant states are colour-coded using the same scheme of figure \ref{fig:1}c. Now, to generate the surrogate data, all the points of $u^{\prime}w^{\prime}$ are replaced by the quantity,
\begin{equation}
u^{\prime}w^{\prime}_{\rm s}=\frac{\sum{u^{\prime}w^{\prime}}}{N_{\rm P}-N_{\rm N}} \times \sign{(u^{\prime}w^{\prime})},
\label{surro_uw}
\end{equation}
where $u^{\prime}w^{\prime}_{\rm s}$ are the constant surrogate values, $N_{\rm P}$ and $N_{\rm N}$ are the number of positive and negative samples, and $\sign$ represents a function which is $+1$ ($u^{\prime}w^{\prime}>0$) or $-1$ ($u^{\prime}w^{\prime}<0$) depending on the argument. Note that if the values of $u^{\prime}w^{\prime}_{\rm s}$ are summed and then divided by the total number of samples, one would simply obtain the time-averaged momentum flux for that 30-min run. Furthermore, if the persistence PDFs of four quadrant states are evaluated from $u^{\prime}w^{\prime}_{\rm s}$ they remain exactly the same as the original ones. Therefore, by employing such surrogate data, momentum transport at any time scale can only be explained through persistence, without focussing on the amplitude variability.

Consequently, if one compares the flux distributions plotted against $t_{p}/\Gamma_{w}$ between the original and surrogate data of $u^{\prime}w^{\prime}$, then any existing difference could be directly related to amplitude variability. Since cumulative distributions are preferred when any two statistical quantities are compared \citep{chamecki2013persistence}, we compute the accumulated values of the fluxes against the persistence time scales, separately for the four quadrants. Thereafter, for each quadrant, we take the absolute difference between the cumulative flux distributions as obtained from the original and surrogate datasets. This is a standard procedure while comparing any two statistical distributions \citep{chowdhuri2021visibility}. Mathematically, such an operation can be expressed as,
\begin{equation}
{\mathcal{L}}_{1} \vert (t_{p}/\Gamma_{w})=|F[u^{\prime}w^{\prime} \vert (t_{p}/\Gamma_{w})]-F[u^{\prime}w^{\prime}_{\rm s} \vert (t_{p}/\Gamma_{w})]|,
\label{l1_norm}
\end{equation}
where the $F$ functions indicate the cumulative flux distributions at a particular scale $t_{p}/\Gamma_{w}$ and ${\mathcal{L}}_{1} \vert (t_{p}/\Gamma_{w})$ is the absolute difference between the original ($u^{\prime}w^{\prime}$) and surrogate ($u^{\prime}w^{\prime}_{\rm s}$) datasets at that scale.

\begin{figure}
\centering
\hspace*{-1.1in}
\includegraphics[width=1.5\textwidth]{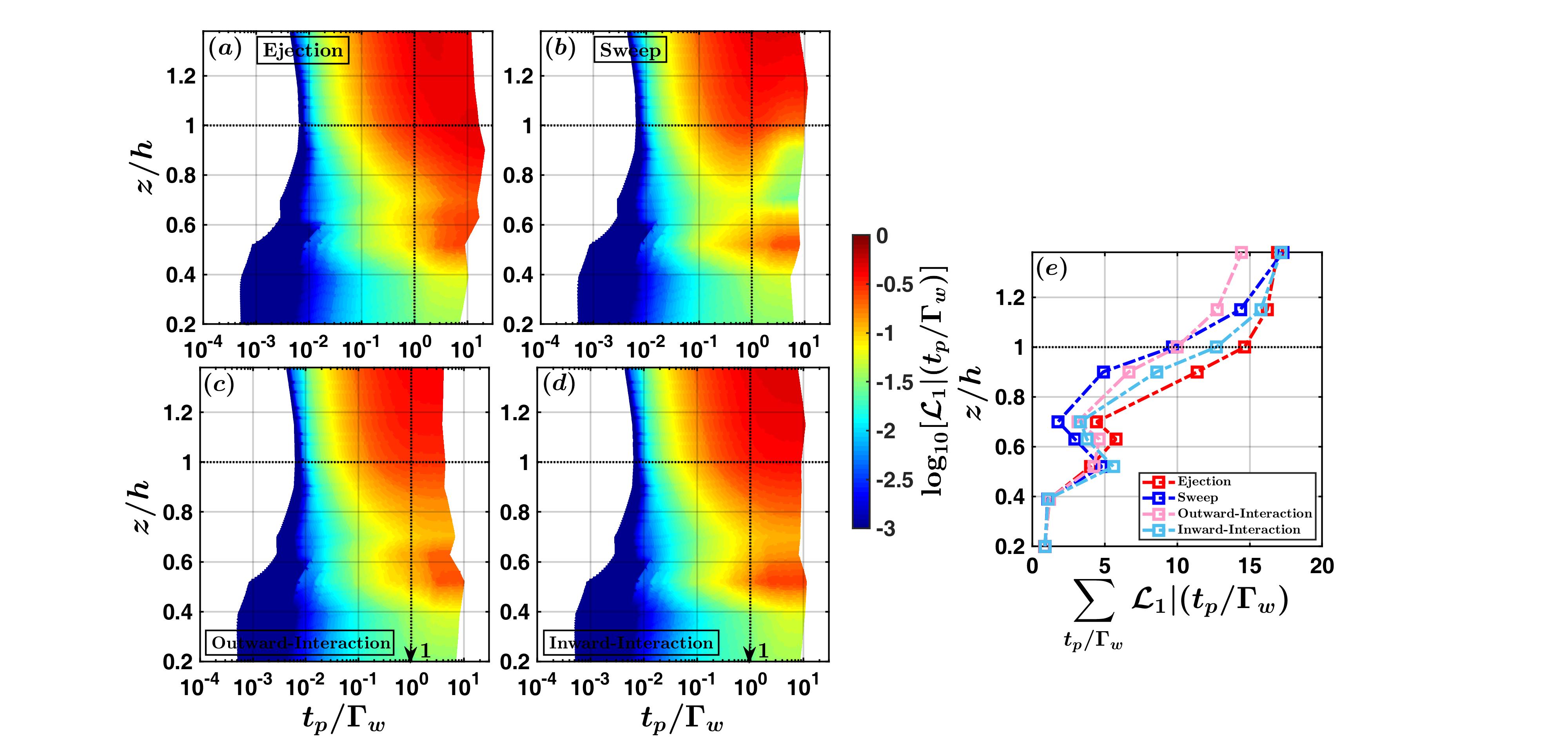}
 \caption{In (a)--(d), quadrant decomposition of the absolute differences between $u^{\prime}w^{\prime}$ distributions (${\mathcal{L}}_{1} \vert (t_{p}/\Gamma_{w})$) are shown against the persistence time scales $t_{p}/\Gamma_{w}$, when evaluated for the original and surrogate time series of $u^{\prime}w^{\prime}$. Note that the original time series preserve both the persistence and amplitude effects, whereas the surrogate ones only contain the persistence effects. For better visualization, the logarithm values of ${\mathcal{L}}_{1} \vert (t_{p}/\Gamma_{w})$ are used. In (e), the vertical profiles of the summed $\mathcal{L}_{1}$ values are shown separately for each quadrant.}
\label{fig:9}
\end{figure}

In figure \ref{fig:9}a--d, we show how ${\mathcal{L}}_{1} \vert (t_{p}/\Gamma_{w})$ vary quadrant-wise when plotted against the persistence time scales $t_{p}/\Gamma_{w}$ and normalized heights $z/h$. For a better visualization, we use the logarithm of ${\mathcal{L}}_{1} \vert (t_{p}/\Gamma_{w})$ while plotting the contours (figure \ref{fig:9}a--d). To highlight the threshold scale $t_{p}/\Gamma_{w}=1$ and $z/h=1$, two black dotted lines (horizontal and vertical) are drawn in figure \ref{fig:9}a--d. The red-shaded contours in figure \ref{fig:9}a--d denote the regions where the largest difference exists between the two cumulative distributions, i.e., where the amplitudes play a stronger role.

By inspecting the contour plots (figure \ref{fig:9}a--d), it is recognized that for the lowest three levels local amplitude effects are negligible across any time scales, as no presence of red-regions can be detected. This indicates, there is almost no temporal variability in the $u^{\prime}w^{\prime}$ time series at heights deep within the canopy. Moreover, since the time-averaged momentum fluxes are itself low at those heights (figure \ref{fig:2}c), the instantaneous flux values remain low too due to the consequence of how the surrogate data are defined in Eq. \ref{surro_uw}. Therefore, despite the organizational structure of the flux events remain same as the other heights (evident from the vertical collapse of persistence PDFs in figure \ref{fig:6}), no contribution is made to the momentum transport at any time scale. Such phenomenon shares a resemblance with the concept of inactive turbulence as proposed by \citet{townsend1961equilibrium} and \citet{bradshaw1967inactive}. 

However, as the heights increase, the range of time scales involved with the red-shaded regions also tend to increase. In fact, at heights $z/h \geq 1$, a broad range of time scales between $0.1 < t_{p}/\Gamma_{w} < 10$ engulf the red regions, indicating a strong presence of amplitudes at such scales. Interestingly, this apparent increase in the range of time scales with height is quite clear for the ejection quadrants as opposed to the rest. For the other three quadrants, a slight break in the red regions is observed between the mid-canopy and above-canopy heights, thereby implying a subtle change in the amplitude effects. 

To further characterize the height variations, in figure \ref{fig:9}e, we show the vertical profiles of $\sum{{\mathcal{L}}_{1} \vert (t_{p}/\Gamma_{w})}$ (summed over all $t_{p}/\Gamma_{w}$ values), separately for the four quadrants (shown as different coloured lines). Although the amplitude effects become increasingly important with $z$, unlike figure \ref{fig:9}a--d, we do not observe any particular dominance from whichever of the four quadrants. Nevertheless, upon closer inspection, the ${\mathcal{L}}_{1}$ values of ejection motions are found to be somewhat greater than the sweeps. Thus, the amplitude effects appear to be a little stronger for the ejections as compared to the sweeps.    

In conclusion, our results from surrogate analysis provide a novel viewpoint by pointing out the range of time scales that are associated with amplitude variability in $u^{\prime}w^{\prime}$ time series. Moreover, we observe that this range is sensitive to the quadrant type and measurement heights. This outcome is compelling enough to make one ponder what type of eddy processes might give rise to such phenomenon. In wall-bounded turbulence, local amplitude variability in either velocity fluctuations or instantaneous momentum flux is usually associated with a modulation effect by the large-scale structures from the outer layer \citep{hutchins2007evidence,baars2015wavelet,gomit2018structure,laskari2018time}. On the other hand, for canopy turbulence, \citet{perret2021stability} have identified a similar presence of amplitude modulation effect but with a change in its mechanism for heights within and above the canopy. To explain this change, they propose a theory that involves forward and backward scattering of turbulent kinetic energy. Quite possibly, our observation showing the range of time scales involved with amplitude variability of $u^{\prime}w^{\prime}$ increases with height, could be related to such mechanisms. However, pursuing that further is beyond the scope of this study.   

\section{Conclusion}
\label{conclusion}
In a dense canopy flow, for the first time, we present a statistical description of the time scales involved with intermittent momentum transport at multiple heights (both within and above the canopy), by using the GoAmazon field-experimental dataset. To compute these time scales, we invoke the concept of persistence, defined as the probability that the local value of a fluctuating field does not change sign up to a certain time. Since the computation of persistence time scales is completely rooted in the temporal domain (unlike Fourier transformation, a non-local decomposition), this analysis procedure provides the unique opportunity to study the time scales of different quadrant events separately. By applying persistence analysis, we seek to answer the research questions posed in the Introduction and in that respect the main findings from our study can be listed as below:
\begin{enumerate}
    \item The probability distributions of persistence time scales ($t_{p}$), associated with four quadrant events, are height-invariant when $t_{p}$ values are normalized by the integral scale of vertical velocity ($\Gamma_{w}$). The collapse of quadrant-wise persistence PDFs under the $\Gamma_{w}$-scaling physically suggests that, the eddy structures whose time scales are comparable to $\Gamma_{w}$, transport momentum both within and above the GoAmazon canopy. However, if the time-averaged fluxes are considered, then at heights deep within the dense canopy $\overline{u^{\prime}w^{\prime}}$ values are approximately 0 due to the presence of the foliage. To resolve this issue, it is proposed that the coherent structures, which reside above the canopy transporting momentum, extend their footprints down to heights as small as $z/h=0.2$. Since $\overline{u^{\prime}w^{\prime}}$ are practically non-existent at those heights, the footprints of the coherent structures do not transport any momentum and therefore they act as inactive motions. On the other hand, if instead of $\Gamma_{w}$, canopy height and friction velocity are used as scaling parameters, then only the time scale distributions of ejections and sweeps collapse while the counter-gradient ones display a clear dependence on height.
    \item The time scales of momentum-transporting events show a power-law signature up to $t_{p}<\Gamma_{w}$, punctuated with an exponential drop corresponding to scales $t_{p} \geq \Gamma_{w}$. From our analysis, we find the exponent of this power-law is nearly equal to $-1.5$. Upon comparing this value with other types of atmospheric and engineering flows, we conclude that such exponents are far from universal owing to their dependence on the surface types over which they are studied.
    \item Based on the statistical features of $u^{\prime}w^{\prime}$ persistence PDFs, two different event classes are recognized. For one of those classes, the time scales of short-lived $u^{\prime}w^{\prime}$ events ($t_{p}<\Gamma_{w}$) follow a power-law behaviour, thereby indicating a presence of statistically self-similar flow structures. On the other hand, the time scales of long-lived events ($t_{p} \geq \Gamma_{w}$) display a clear deviation from the power-law, which can be associated with the existence of large-scale coherent structures. It is discovered that around 80\% of the momentum is transported through the long-lived events at heights immediately above the canopy while the short-lived ones only contribute marginally ($\approx$ 20\%). 
    \item To explore any connection between persistence and amplitude variability while evaluating the flux contributions from $u^{\prime}w^{\prime}$ events, a surrogate dataset is used. By employing this dataset, we establish that a wide range of persistence time scales are involved with amplitude variations in $u^{\prime}w^{\prime}$ as one transitions from within to above the canopy. In fact, at heights $z/h \geq 1$, the events, having time scales in the range $0.1 < t_{p}/\Gamma_{w} < 10$, are associated with strong variations in $u^{\prime}w^{\prime}$ values. However, for mid canopy heights, this range is mostly restricted between $1 < t_{p}/\Gamma_{w} < 10$. Therefore, as the heights approach the canopy top, the strong amplitude effects pervade to time scales which are much smaller than $\Gamma_{w}$ at least by an order of magnitude. We hypothesize that such height-dependent relationships between persistence time scales and variability in instantaneous $u^{\prime}w^{\prime}$ values could be the consequence of an amplitude modulation effect as proposed by \citet{perret2021stability} for canopy flows.    
\end{enumerate}

In summary, the above results are quite useful to deduce the nuances of intermittent momentum transport in dense canopy flows and hopefully pave the way towards the development of better parameterization schemes, accounting for the intermittency aspect. Currently, we only consider near-neutral conditions, but in future more stratification regimes (highly-convective to stably-stratified) will be surveyed, to assess the varying effects of shear and buoyancy on the time scales of intermittent flux transports. A limitation of the present study is the lack of three-dimensional information due to the usage of point-wise temporal measurements that only span vertically up to nine heights. Therefore, we can not explicitly examine the connection between the spatial geometry of coherent structures and persistence time scales. However, in our future efforts, large-eddy simulation results of a dense canopy flow will be analyzed to shed light on this aspect.  

\backsection[Acknowledgements]{The U.S. Department of Energy supported the field studies as part of the GoAmazon project (grant SC0011075). Fundação de Amparo à Pesquisa do Estado de São Paulo (FAPESP) and Fundação de Amparo à Pesquisa do Estado do Amazonas (FAPEAM) funded the Brazilian component of the field studies. The Large scale Biosphere-Atmosphere Experiment in Amazonia (LBA) provided logistic support and made the flux tower and housing unit available to complete the field studies. The authors thank the three anonymous reviewers for their helpful comments. SC also thanks Dr. Giovanni Iacobello for providing the MatLab code to compute the ${\mathcal{L}}_{1}$-norm. }

\backsection[Funding]{SC gratefully acknowledges the support from Indian Institute of Tropical Meteorology (IITM), an autonomous institute fully funded by the Ministry of Earth Sciences, Government of India. TB acknowledges the funding support from the University of California Office of the President (UCOP) grant LFR-20-653572 (UC Lab-Fees); the National Science Foundation (NSF) grants NSF-AGS-PDM-2146520 (CAREER), NSF-OISE-2114740 (AccelNet) and NSF-EAR-2052581 (RAPID); the United States Department of Agriculture (USDA) grant 2021-67022-35908 (NIFA); and a cost reimbursable agreement with the USDA Forest Service 20-CR-11242306-072.}

\backsection[Declaration of interests]{The authors report no conflict of interest.}

\backsection[Author ORCID]{S. Chowdhuri, 0000-0002-5518-7701; K. Ghannam, 0000-0002-2542-6388; T. Banerjee, 0000-0002-5153-9474}

\backsection[Author contributions]{All the authors designed the study and SC carried out the analyses. SC prepared the figures and wrote the initial draft of the manuscript. KG and TB provided their corrections, comments, and suggestions. All the authors read the final draft and agreed to all the changes.}

\appendix
\begin{appendices}

\section{Polar quadrant analysis}
\label{app_B}

\begin{figure}
\centering
\hspace*{-0.5in}
\includegraphics[width=1.5\textwidth]{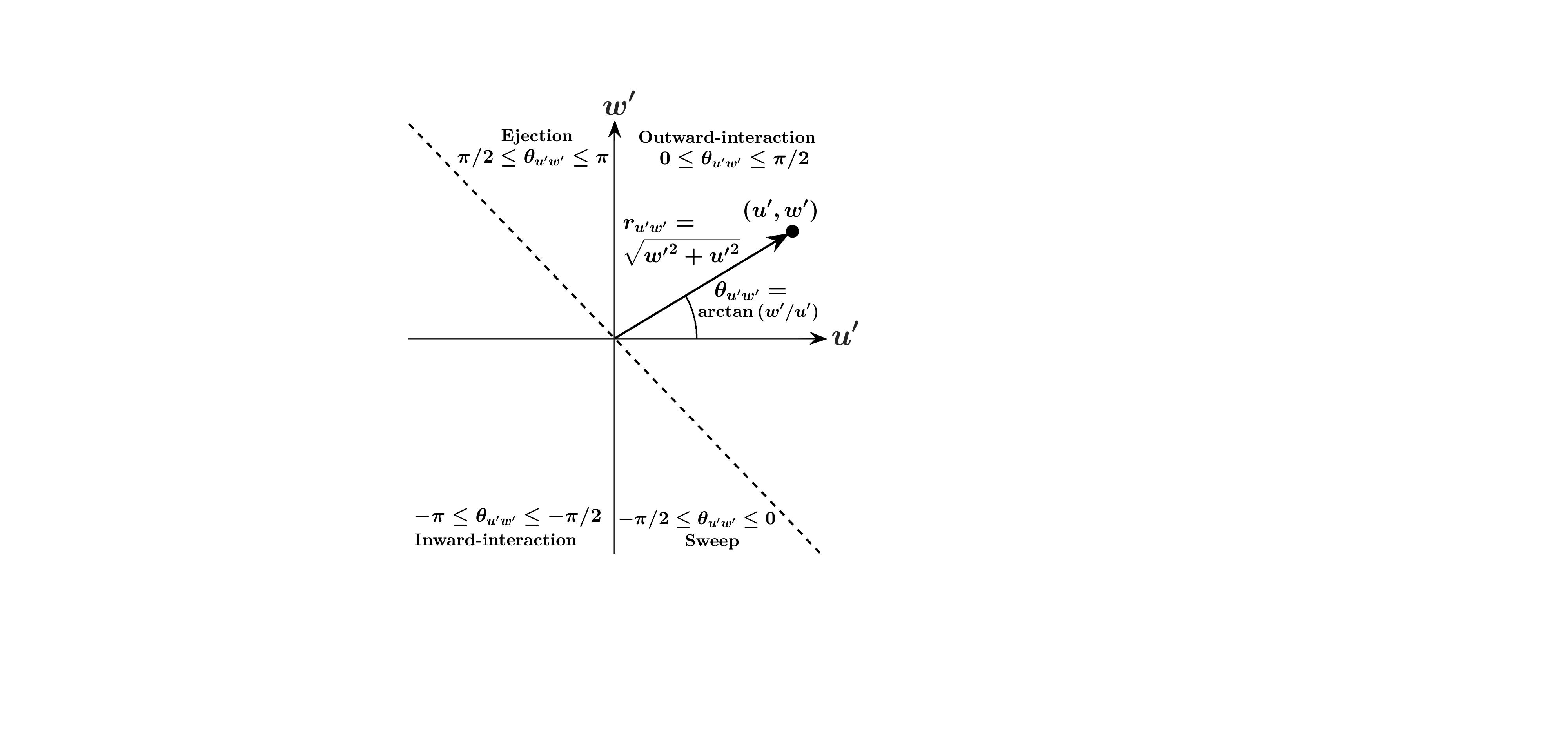}
 \caption{An example is shown to explain the concepts of polar-quadrant analysis. A momentum flux event can be described through two quantities, which are amplitude $r_{u^{\prime}w^{\prime}}$ and phase angle $\theta_{u^{\prime}w^{\prime}}$. Moreover, each of the four quadrants can be identified through the ranges in the phase angle $\theta_{u^{\prime}w^{\prime}}$.}
\label{fig:A2}
\end{figure}

As opposed to the quadrant analysis in Cartesian co-ordinates, in polar-quadrant analysis the quadrant plane is envisaged as a phase space, where each point is designated with two parameters such as the phase angles and amplitudes \citep{mahrt1984heat,chowdhuri2020persistence_b}. In Cartesian quadrant analysis, the coupling between the $u^{\prime}$ and $w^{\prime}$ signals is usually studied through their joint probability density functions (JPDFs), plotted for a particular height \citep[e.g.][]{li2011coherent,chamecki2013persistence,wallace2016quadrant}. However, a systematic investigation of changes in the nature of coupling between $u^{\prime}$ and $w^{\prime}$ with increasing heights would require a visual inspection of $u^{\prime}$-$w^{\prime}$ JPDFs at all such heights. This procedure becomes cumbersome when multiple heights are involved. As compared to the Cartesian analysis, a distinct advantage of polar-quadrant analysis is it allows to inspect the height variations in quadrant plots quite easily.

Figure \ref{fig:A2} graphically illustrates the concept of $u^{\prime}$-$w^{\prime}$ quadrant plane from the perspective of a polar reference frame. As mentioned before, each point on the $u^{\prime}$-$w^{\prime}$ quadrant plane can be associated with an amplitude $r_{u^{\prime}w^{\prime}}$ and phase angle $\theta_{u^{\prime}w^{\prime}}$, expressed as,
\begin{align}
\theta_{u^{\prime}w^{\prime}}=\arctan{(w^{\prime}/u^{\prime})}\\
r_{u^{\prime}w^{\prime}}=\sqrt{{w^{\prime}}^2+{u^{\prime}}^2}
\label{flux_vector}.
\end{align}
The values of $\theta_{u^{\prime}w^{\prime}}$ vary between $-\pi$ to $\pi$ and their ranges are related to the four different quadrants as demonstrated in table \ref{tab:1}. 

\begin{table}
  \begin{center}
\def~{\hphantom{0}}
  \begin{tabular}{lccc}
Phase angle ($\theta_{u^{\prime}w^{\prime}}$) & $u^{\prime}$-$w^{\prime}$ plane & Quadrant type & Quadrant name \\
\hline
\\
$0 \leq \theta_{u^{\prime}w^{\prime}} \leq \pi/2$ & $u^{\prime}>$ 0, $w^{\prime}>$ 0 & Counter-gradient & Outward-interaction (OI)\\
\\
$\pi/2 \leq \theta_{u^{\prime}w^{\prime}} \leq \pi$ & $u^{\prime}<$ 0, $w^{\prime}>$ 0 & Co-gradient & Ejection (E)\\
\\
$-\pi \leq \theta_{u^{\prime}w^{\prime}} \leq -\pi/2$ & $u^{\prime}<$ 0, $w^{\prime}<$ 0 & Counter-gradient &  Inward-interaction (II)\\
\\
$-\pi/2 \leq \theta_{u^{\prime}w^{\prime}} \leq 0$ & $u^{\prime}>$ 0, $w^{\prime}<$ 0 & Co-gradient & Sweep (S)\\
  \end{tabular}
    \caption{Various parameters to describe the four quadrants of $u^{\prime}$-$w^{\prime}$}
 \label{tab:1}
  \end{center}
\end{table}

In the polar co-ordinate system, the instantaneous momentum flux $u^{\prime}w^{\prime}$ associated with each point is expressed as,
\begin{equation}
u^{\prime}w^{\prime}=r_{u^{\prime}w^{\prime}} \cos{(\theta_{u^{\prime}w^{\prime}})} \times r_{u^{\prime}w^{\prime}} \sin{(\theta_{u^{\prime}w^{\prime}})} \implies \frac{1}{2} r_{u^{\prime}w^{\prime}}^{2} \sin{(2 \theta_{u^{\prime}w^{\prime}})}.
\label{flux_value_pre}
\end{equation}
In Eq. (\ref{flux_value_pre}), since $r_{u^{\prime}w^{\prime}}^{2}$ is a positive definite quantity, the distribution of $u^{\prime}w^{\prime}$ fluxes among the four quadrants is primarily decided by $\theta_{u^{\prime}w^{\prime}}$. On this note, the PDFs of $\theta_{u^{\prime}w^{\prime}}$ ($P(\theta_{u^{\prime}w^{\prime}})$) are related to the fraction of time spent by the signal in a particular quadrant state. This is because,
\begin{equation}
T_{f\rm X}=\int_{-\pi}^{\pi} P(\theta_{u^{\prime}w^{\prime}}) I_{\rm X}(\theta_{u^{\prime}w^{\prime}}) \ d\theta_{u^{\prime}w^{\prime}},
\label{T_f}
\end{equation}
where $T_{f\rm X}$ is time-fraction spent in quadrant X (X could be any one of the four quadrants) and $I_{\rm X}(\theta_{u^{\prime}w^{\prime}})$ is the identity function that is unity when $\theta_{u^{\prime}w^{\prime}}$ lies within quadrant X or zero elsewise. To compute $P(\theta_{u^{\prime}w^{\prime}})$ from the present data at hand, we use 60 bins of $\theta_{u^{\prime}w^{\prime}}$, spanning between $-\pi$ to $\pi$ \citep{chowdhuri2020persistence_b}. Note that for each 30-min run, the total possible values of $\theta_{u^{\prime}w^{\prime}}$ are equal to the number of samples, i.e., 36000. Since the PDFs $P(\theta_{u^{\prime}w^{\prime}})$ are averaged over 93 half-hour blocks of near-neutral runs, this results in close to half-a-million number of samples when all the ensemble members are considered together. 

In addition to $P(\theta_{u^{\prime}w^{\prime}})$, the $u^{\prime}w^{\prime}$ fluxes associated with $\theta_{u^{\prime}w^{\prime}}$ can be computed as,
\begin{equation}
\langle u^{\prime}w^{\prime} \vert \{ \theta_{u^{\prime}w^{\prime}}(i)<\theta_{u^{\prime}w^{\prime}}<{\theta_{u^{\prime}w^{\prime}}(i)+d\theta_{u^{\prime}w^{\prime}}\}} \rangle=\frac{\sum u^{\prime}w^{\prime}(i)} {N \times d\theta_{u^{\prime}w^{\prime}}},
\label{Flux_theta}
\end{equation}
where $i$ is the bin index, $d\theta_{u^{\prime}w^{\prime}}$ is the bin width, and $N$ is the total number of samples in a 30-min run (36000 in our case). The division by $N$ and $d\theta_{u^{\prime}w^{\prime}}$ is to ensure that when integrated over $\theta_{u^{\prime}w^{\prime}}$, it would yield $\overline{u^{\prime}w^{\prime}}$. For brevity, the quantity at the left-hand-side of Eq. (\ref{Flux_theta}), is denoted as $\langle u^{\prime}w^{\prime} \vert (\theta_{u^{\prime}w^{\prime}})\rangle$. Another important aspect is, if in the ejection or sweep quadrant, $\theta_{u^{\prime}w^{\prime}}$ lies to the left or right of the black dashed line ($w^{\prime}=-u^{\prime}$, figure \ref{fig:A2}), then we can infer that $-u^{\prime}w^{\prime}$ fluxes associated with those phase angles (see Eq. (\ref{Flux_theta})) are dominated by the magnitudes of $u^{\prime}$ rather than of $w^{\prime}$.

\section{$u$, $w$ spectra and $u$-$w$ cospectra}
\label{app_C}

\begin{figure}
\centering
\hspace*{-1.6in}
\includegraphics[width=1.5\textwidth]{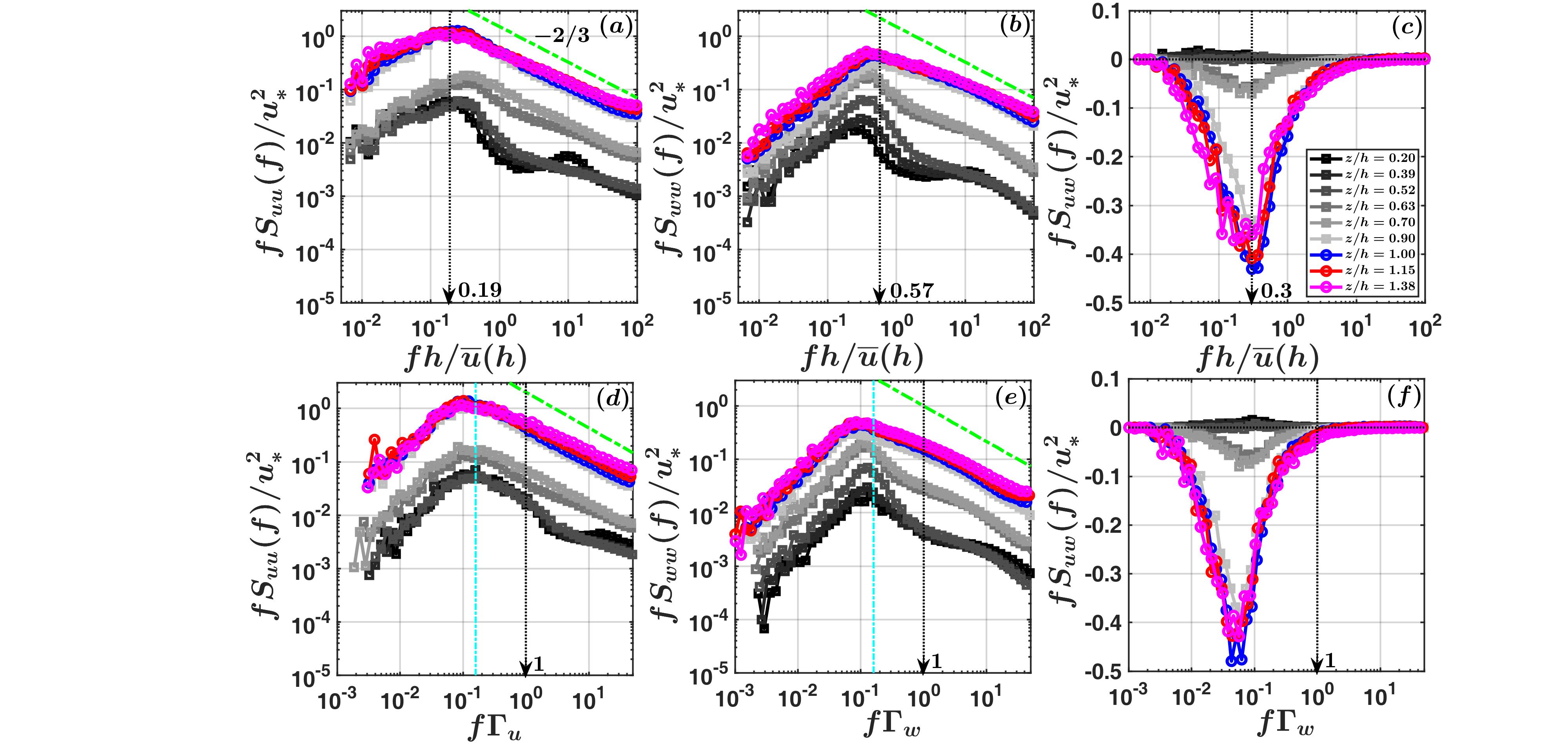}
\caption{The ensemble averaged (a, d) $u$, (b, e) $w$ spectra and (c, f) $u$-$w$ cospectra are shown for the near-neutral runs, where the spectral amplitudes are normalized by $u_{*}^2$ but the frequencies are scaled in two different ways. The spectral frequencies in (a), (b), and (c) are scaled by $h$ and $\overline{u}(h)$ (mean velocity at canopy top), whereas in the lower panels (d, e, and f) the integral time scales are used, such as: $\Gamma_{u}$ and $\Gamma_{w}$.}
\label{fig:A3}
\end{figure}

In this appendix, we report the ensemble-averaged $u$, $w$ spectra and $u$-$w$ cospectra from the near-neutral conditions, corresponding to the heights within and above the canopy (figure \ref{fig:A3}). These results highlight how the spectra and cospectra compare with the previous studies and whether the $u$-$w$ cospectra complement the information obtained from persistence analysis of $u^{\prime}w^{\prime}$ signal. Note that while plotting the spectra and cospectra, the frequencies ($f$) are scaled in two ways, first by $h$ and $\overline{u}(h)$ (canopy height and the mean velocity at canopy top) and second by the Eulerian integral timescales ($\Gamma_{x}$, where $x=u,w$). Although two different scalings are used for $f$, the premultiplied spectral and cospectral amplitudes ($fS_{xx}(f)$, where $x=u,w$, and $fS_{uw}(f)$) for both cases are scaled by $u^2_{*}$, where $u_{*}$ is the friction velocity at the canopy top. The scaling, $f h/\overline{u}(h)$, is to ensure that the peak frequency values of $u$, $w$ spectra and $u$-$w$ cospectra can be compared with the other experimental results of \citet{dupont2012influence} and \citet{su1998turbulent}, respectively (figure \ref{fig:A3}a--c). On the other hand, the scaling of $f$ by $\Gamma_{x}$ is to provide a context while comparing the results obtained from persistence analysis with the more conventional spectral methods (figure \ref{fig:A3}d--f).

As one may note, in both figure \ref{fig:A3}a--b and d--e irrespective of the scalings used, a well-defined inertial subrange with $-2/3$ slope can be observed in the $u$, $w$ spectra at all the heights except the lowest three ones. In figure \ref{fig:A3}a--c, for heights $z/h \geq 1$, the spectral and cospectral amplitudes attain their peaks at those scaled frequency values which match exceptionally well with the previous studies of \citet{dupont2012influence} and \citet{su1998turbulent} from near-neutral conditions. To illustrate this point, the peak values of $f h/\overline{u}(h)$ from \citet{dupont2012influence} and \citet{su1998turbulent} are marked as vertical black dashed lines in figure \ref{fig:A3}a--c. 

In contrary to figure \ref{fig:A3}a--c, the vertical black dashed lines in figure \ref{fig:A3}d--f represent the threshold $f\Gamma_{x}=1$, which are used to demarcate the periods larger or smaller than the integral scales. From figure \ref{fig:A3}d--f one can see that the peaks of spectral and cospectral amplitudes are located around those periods which are larger than the integral scales. Particularly, in accordance with \citet{kaimal1994atmospheric}, the spectral amplitudes of both $u$ and $w$ attain their peaks at $f\Gamma_{x}=1/(2\pi)$ for all the nine measurement heights (cyan dashed lines in figure \ref{fig:A3}d--e). By focussing on $u$-$w$ cospectra in figure \ref{fig:A3}f, a qualitative resemblance can be noted with figure \ref{fig:7}f, where we presented the flux contributions from short- and long-lived events separately. In figure \ref{fig:7}f, we found that almost 80\% of the streamwise momentum fluxes are carried by those persistence patterns whose time scales are larger than the integral scale of vertical velocity. This finding agrees quite well with the result obtained from $u$-$w$ cospectra in figure \ref{fig:A3}f, where the periods larger than $\Gamma_{w}$ mostly contribute to the $u^{\prime}w^{\prime}$ flux.  

\end{appendices}

\bibliographystyle{jfm}
\bibliography{Intermittent_momentum.bib}
\end{document}